        \documentclass[structabstract]{aa}  
\usepackage{natbib,twoopt}
\usepackage[breaklinks=true]{hyperref} %% to avoid \citeads line fills
\bibpunct{(}{)}{;}{a}{}{,}             %% natbib format for A&A and ApJ
\makeatletter
  \newcommandtwoopt{\citeads}[3][][]{\href{http://adsabs.harvard.edu/abs/#3}%
    {\def\hyper@linkstart##1##2{}%
     \let\hyper@linkend\@empty\citealp[#1][#2]{#3}}}
  \newcommandtwoopt{\citepads}[3][][]{\href{http://adsabs.harvard.edu/abs/#3}%
    {\def\hyper@linkstart##1##2{}%
     \let\hyper@linkend\@empty\citep[#1][#2]{#3}}}
  \newcommandtwoopt{\citetads}[3][][]{\href{http://adsabs.harvard.edu/abs/#3}%
    {\def\hyper@linkstart##1##2{}%
     \let\hyper@linkend\@empty\citet[#1][#2]{#3}}}
  \newcommandtwoopt{\citeyearads}[3][][]%
    {\href{http://adsabs.harvard.edu/abs/#3}
    {\def\hyper@linkstart##1##2{}%
     \let\hyper@linkend\@empty\citeyear[#1][#2]{#3}}}
\makeatother

\usepackage{graphicx}
\usepackage{units}
\usepackage{amsmath}
\usepackage{subfigure}
\usepackage{txfonts}
\begin{document}
   \title{Porous dust grains in debris disks}

   %\subtitle{}

   \author{Florian Kirchschlager
         % \inst{1}
          \and
           Sebastian Wolf%\inst{1}
          }

   \institute{Institut f\"ur Theoretische Physik und Astrophysik, Christian-Albrechts-Universit\"at zu Kiel,
              Leibnizstra\ss e 15, 24098 Kiel, Germany\\
              \email{kirchschlager@astrophysik.uni-kiel.de\\
\phantom{email:}wolf@astrophysik.uni-kiel.de}
             }

   \date{Received 2 October 2012; accepted 25 January 2013}

% \abstract{}{}{}{}{} 
% 5 {} token are mandatory
 
%   \abstract{
% %\newline~ a \newline~ b \newline~ c 
% }
  \abstract
   {When modeling the density and grain size distribution in debris disks, the minimum particle size is often significantly larger than the corresponding blowout size. While the dust particles are usually modeled as compact, homogenous spheres, we instead investigate the impact of porosity.}
   {The optical properties of porous particles are determined, and the influences of porosity on the blowout size and dust temperatures investigated.}
   {Using the method of discrete dipole approximation, we calculate the scattering and absorption cross sections of porous particles and derive the blowout size and the behavior of the dust temperature.}
   {We determine the efficiency factors of absorption and the radiation pressure for porous particles, and investigate the influence on the $\beta$-ratio. Blowout sizes are calculated for various stellar luminosities and porosities, and an approximation equation is derived to estimate the blowout size as a function of these parameters. Furthermore, we investigate the influence of the porosity on the dust equilibrium temperature.}
   {The blowout size increases with the particle porosity and stellar luminosity. The dust temperature of porous particles is lower than  the one of the compact spheres, in particular the temperature of blowout grains decreases for porous particles.}
  % {huhuhuh dfsa sdf} % leave it empty if necessary  
  % {\vspace{3.9cm}}


% \titlerunning{Porous dust grains}
\maketitle  
%#########################################################################################################################################
%#########################################################################################################################################
%#########################################################################################################################################
%#########################################################################################################################################
%#########################################################################################################################################
%#########################################################################################################################################
%###########################################            1        #########################################################################
%#########################################################################################################################################
%#########################################################################################################################################
%#########################################################################################################################################
\section{Introduction}
Debris disks contain micrometer-sized dust grains, which can be observed in  scattered light in the optical to near-infrared wavelength range and by thermal emission at longer wavelengths. To interpret observations of debris disks these particles are often assumed to be compact, homogenous, and spherical (e.g \citealt{Roccatagliata2009}, \citealt{Ertel2011, Ertel2012b}). This is a convenient simplification, since it makes the Mie scattering formalism applicable (\citealt{Mie}, \citealt{WolfVoshchinnikov04}). However, there are various processes that would tend to create more complex, e.g. inhomogeneous or almost arbitrarily shaped, particles, instead. In cold clouds gas  molecules can  freeze out on the surface of dust grains, so that the particles  are composed of multilayer, inhomogeneous materials (\citealt{Kruegel}). In young, gas-rich circumstellar disks, collisions of particles occur in which the dust grains either agglomerate and  form larger particles  or  get shattered in fragments with complex shapes. In this context, porosity might support the grain growth process in the disk (\citealt{BlumWurm2008}).

Modeling recent observations of dust in various environments also motivates the concept of porous dust grains. For example, \cite{vanBreemen2011} used a distribution of hollow spheres to fit the $\unit[9.7]{\mu m}$ silicate absorption profile of the ISM based on the model of \cite{Min2005}. \cite{Demyk1999} applied some degree of porosity, which was necessary to reproduce silicate features in the spectrum of dust around the protostellar objects RAFGL7009S and IRAS 19110+1045. Scattered light images of the circumstellar disk surrounding the T Tauri star IM Lupi may indicate the presence of fluffy aggregates (\citealt{Pinte2008}). \cite{Birnstiel2010} and \cite{Ricci2010,Ricci2012} used porous spherical grains to specify the dust opacity in a sample of young protoplanetary disks. For a general study of the influence of porosity on the dust opacity in protoplanetary disks, see \cite{Semenov2003}. 

Modeling debris disks is an important part of understanding the evolution and formation of stellar disks, hence the evolution of planetary systems. With that in mind, \cite{Augereau1999, Augereau2001}, \cite{Wyatt2002}, and \cite{Fitzgerald2007} used porous grains for disk modeling, resulting in suitable approaches to reproduce observational data of  HR$\,$4796$\,$A, $\beta\,$Pictoris, Fomalhaut, and AU Microscopii, respectively. \cite{Acke2012} found evidence of fluffy aggregates in the disk of Fomalhaut. These are required for the thermal emission characteristic of small grains and the scattering behavior of large particles. A dependence exists between grain porosity and dust temperature (\citealt{Greenberg1990}, \citealt{Vosh2}), causing a shift in the infrared emission to longer wavelengths. 

Our study is motivated by the fact that debris disk modeling using compact grains often reveals 
a significant discrepancy between  the minimum particle size and the calculated blowout size (e.g. \citealt{Roccatagliata2009}, \citealt{Ertel2011}). The influence of porosity on the radiation pressure has already been considered  in previous studies. \citet{Mukai92} and \citet{Kimura1997}  examined the $\beta$-ratio for fractal aggregates using effective medium theory, while \citet{Koehler2007} applied  discrete dipole approximation. We investigate porous particles and outline various aspects as the absorption efficiency factor, the blowout size, and dust temperatures. In  Sect.  \ref{losgehts} we provide an overview of the applied numerical tool for calculating the efficiency factors and describe our model of porous dust grains. The analysis of the absorption cross sections is presented in Sect. \ref{firstres}, and the blowout size as a function of selected parameters is discussed in Sect. \ref{Kapitel4}. Finally, we present the equilibrium temperature of porous dust particles in Sect. \ref{Staubtemp}. 

%#########################################################################################################################################
%#########################################################################################################################################
%#########################################################################################################################################
%#########################################################################################################################################
%#########################################################################################################################################
%#########################################################################################################################################
%###########################################            2        #########################################################################
%#########################################################################################################################################
%#########################################################################################################################################
%#########################################################################################################################################
\section{Calculation methods}
\label{losgehts}
\subsection{Porous grain model}
\label{AbschnittQuerschnitt}
Within the infinite variety of porous particles we consider those that are spherical and from which material is removed so that voids are formed. The advantage of this model is the small number of parameters describing each particle. A debris disk contains a vast number of grains of different shapes, sizes, and orientations. The appearance of the disk results from averaged optical properties. This motivates  for adopting a grain model that results from an ensemble of particles with similar geometric  properties. The spherical grain model with voids satisfies this task, if  the voids are distributed uniformly over the  volume and thus the grain orientation is negligible.

Porosity is the ratio of the vacuum volume to the total volume of the grain,
\begin{align}
 \mathcal{P}&=V_{\textup{vacuum}}/V_{\textup{total}}=1-V_{\textup{solid}}/V_{\textup{total}},\\
  \mathcal{P}&\in \left[0,1\right].\nonumber
\end{align}
For example, a compact sphere is given for $ \mathcal{P} = 0 $. The radius $ a $ is denoted as the minimum radius of a sphere that encases the grain.

Two grains with equal radius and porosity  can have fundamentally different hole sizes, considering the examples of a hollow sphere and a grain that contains many small holes. In principle,  holes of different sizes can be assumed  within a grain. \citet{Vosh1} adopted  a size distribution of the holes in their model, in which the  probability of the presence of a hole is assumed to be inversely proportional to its volume. The holes in our grain model are determined both by the number of dipoles and the process of the grain generation (Sect. \ref{genera}).

%#########################################################################################################################################
%#########################################################################################################################################
%#########################################################################################################################################
%#########################################################################################################################################
%#########################################################################################################################################
%#########################################################################################################################################
%###########################################            2.2        #######################################################################
%#########################################################################################################################################
%#########################################################################################################################################
%#########################################################################################################################################
\subsection{DDSCAT}
\label{Vollkugeltests}
We used the program DDSCAT to calculate the scattering and absorption cross sections of  irregular particles (\citealt{Draine1994,Draine2010}). It is based  on the theory of discrete dipole approximation  (DDA), where a continuous particle is approximated by the corresponding spatial distribution of  $N$ discrete polarizable points on a cubic lattice. The great advantage of the DDA is the high flexibility of the particle structure, but the applicability is limited by the conditions that the lattice constant $d_{\textup{l}}$ must be small compared to the wavelength and to the scales of the internal grain structure. The first condition is roughly expressed by (\citealt{DraineGood}, \citealt{Draine1994} and \citealt{Draine2000})
\begin{align}
 |n|kd_{\textup{l}}<1,& \label{Kriterium}
\end{align}
where $n$ is the complex refractive index and $k=2\pi/ \lambda$ the circular wavenumber. This criterion is satisfied for  $|n-1|\lesssim3$ (\citealt{Draine2010}).  With the volume of the material in the particle, $V=N(d_{\textup{l}})^3=4/3\pi a^3(1-\mathcal{P})$, it follows $a=\left(\frac{3}{4\pi}\frac{N}{(1-\mathcal{P})}\right)^{1/3}d_{\textup{l}}=\left(\frac{3}{4\pi}\frac{N}{(1-\mathcal{P})}\right)^{1/3}  \frac{|n|kd_{\textup{l}}}{|n|} \frac{\lambda}{2\pi} $, and thus with condition \ref{Kriterium}
\begin{align}
a< 0.09873\frac{\lambda}{|n|}\left(\frac{N}{1-\mathcal{P}}\right)^{1/3}.\label{gultig}
\end{align}
For our studies we use astronomical silicate (\citealt{Draine2003a,Draine2003b}) with a bulk density of $\rho=\unit[3.5]{g\,cm^{-3}}$ (\citealt{Draine2003}).

Before applying the DDSCAT to investigate the behavior of porous grains, we verified the optimum parameter settings (number of dipoles used, etc.) in the case of compact spherical grains. Therefore, we calculate the absorption cross section with the DDSCAT and compare the results with absorption cross sections i) calculated  via Mie theory and ii) using the DDSCAT  with a number of dipoles decreased by a factor of 2. The results are summarized in Appendix \ref{review}.

In Fig. \ref{gultigBild} the applicability of the DDSCAT is illustrated for a compact spherical grain, represented by $\sim$500,000 dipoles. According to  condition \ref{gultig}, the DDSCAT delivers suitable results for the area  below the solid line. While for the solid line the refractive index of  the silicate is taken into account, the crosses are derived by the tests with the compact grains (Appendix \ref{review}). These crosses mark the used upper limit of applicability.
\begin{figure}[t!]
	\centering
\includegraphics[width=1.0\linewidth]{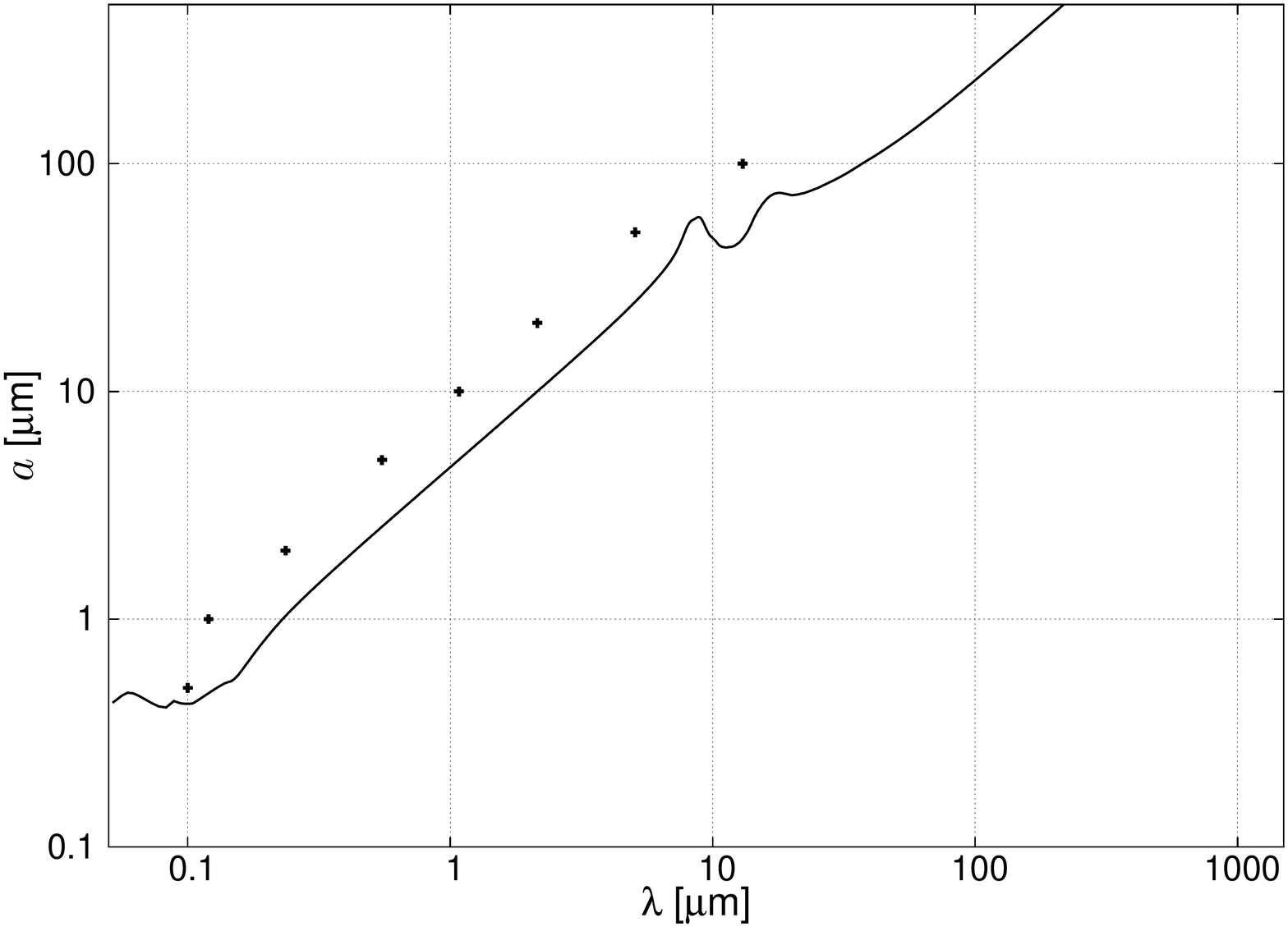}
\caption{The limit of the applicability of the DDSCAT for a compact spherical grain of  astronomical silicate, represented by $\sim$500,000 dipoles. 
According to condition \ref{gultig}, the DDSCAT delivers suitable results for particles with radius $a< 0.09873\frac{\lambda}{|n|}\left(\frac{N}{1-\mathcal{P}}\right)^{1/3}$ (solid line).
In Appendix \ref{review} tests with the DDSCAT are presented, and we define the crosses as the upper limit.}
\label{gultigBild}
\end{figure}
%#########################################################################################################################################
%#########################################################################################################################################
%#########################################################################################################################################
%#########################################################################################################################################
%#########################################################################################################################################
%#########################################################################################################################################
%###########################################            2.3        #######################################################################
%#########################################################################################################################################
%#########################################################################################################################################
%#########################################################################################################################################
\subsection{Generating porous grains}
\label{genera}
\begin{figure}[t!]
	\centering
		\subfigure{\includegraphics[width=0.3\linewidth]{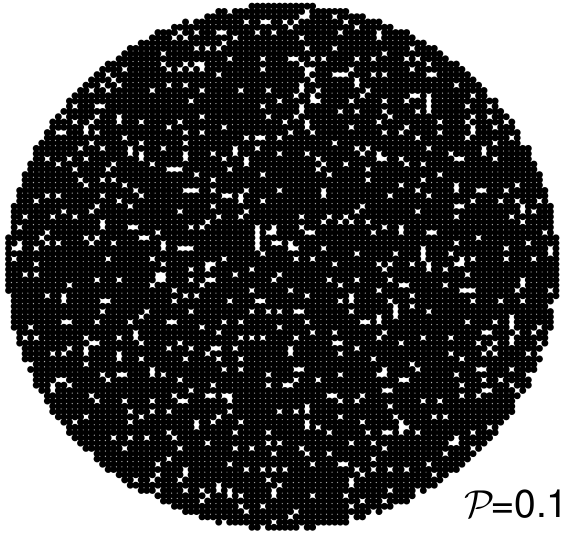}}
		\subfigure{\includegraphics[width=0.3\linewidth]{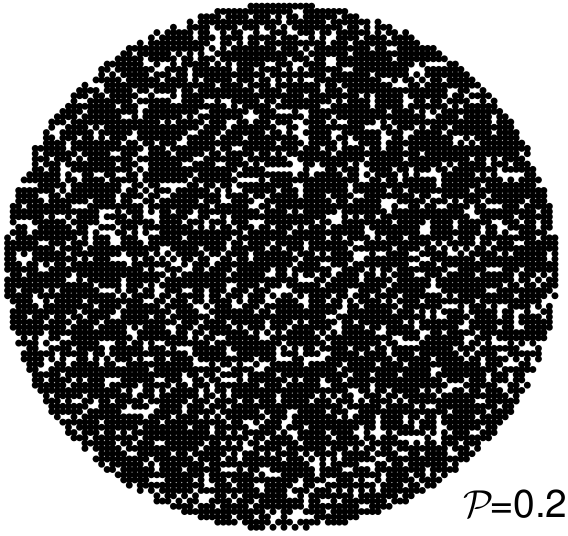}}
		\subfigure{\includegraphics[width=0.3\linewidth]{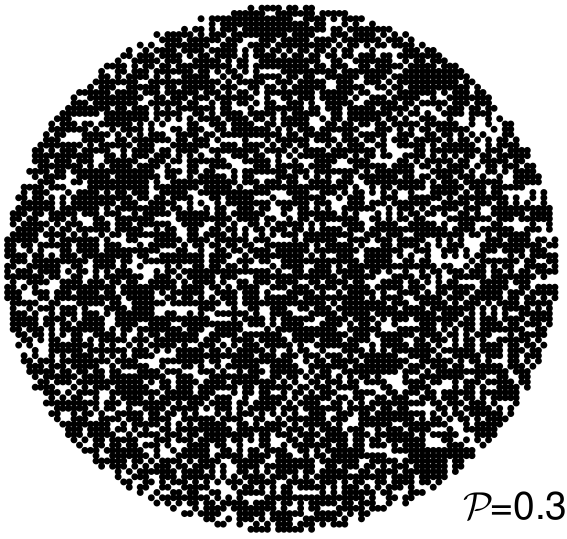}}
		\subfigure{\includegraphics[width=0.3\linewidth]{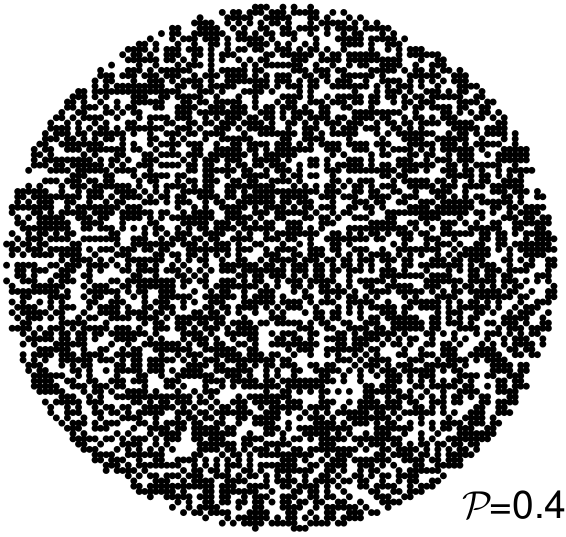}}
		\subfigure{\includegraphics[width=0.3\linewidth]{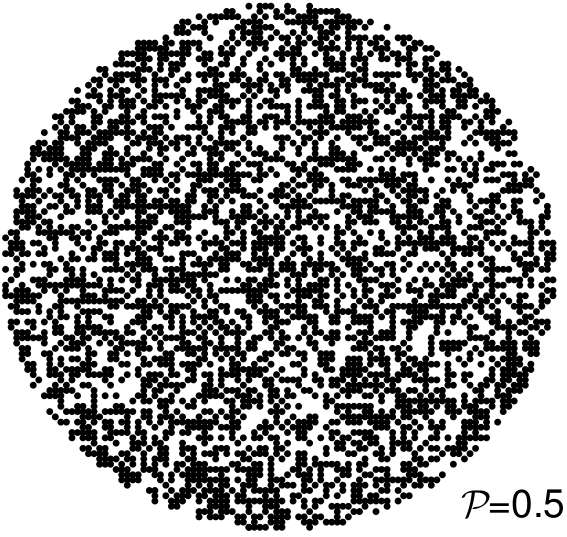}}
		\subfigure{\includegraphics[width=0.3\linewidth]{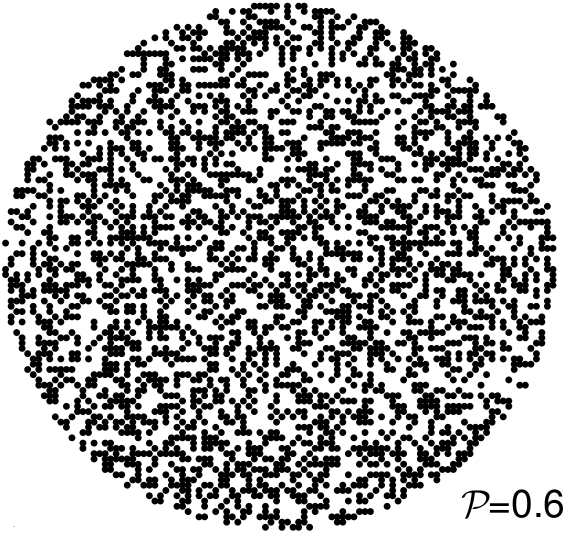}}
	\caption{Particle cross sections through the middle of the grain for porosities from $\mathcal{P}=0.1$ to $\mathcal{P}=0.6$.}
	\label{Teilchenquerschnitt}
\end{figure}

We briefly describe the process of generating porous model grains: $N'=519,832$ dipoles are arranged on a cubic lattice, forming the shape of a compact sphere. A random number generator is used to remove single dipoles to create grains with a porosity $\mathcal{P}$. The remaining dipole arrangement contains $ N=(1-\mathcal{P})\cdot N'$ dipoles, forming a sphere with voids. Each dipole represents a cubic subvolume  $(d_{\textup{l}})^3$ of the  surrounding material, while a removed dipole represents a cubic void of the same size. The large number of dipoles and the application of the random generator ensure a statistical  distribution of the holes. Thus, no preferential direction exists and the optical properties  are roughly independent of the grain orientation. The number of dipoles for  the particles  used in this paper are listed in Table \ref{sovieleNs}. The particle cross sections are shown for different porosities in Fig. \ref{Teilchenquerschnitt}. Some dipoles seem to be  detached from the rest of the complex, but it should be considered that these dipoles may be connected to the particle in the direction perpendicular to the presented cross section. Nevertheless, the procedure of particle generation does not prevent the occurrence of single dipoles, which are isolated from the  rest of the configuration. For $\mathcal{P}=0.1$ the statistical number of  isolated dipoles is negligible  if compared to the number of dipoles of the entire grain, but this number rises sharply with porosity. Therefore, only particles with porosities up to $\mathcal{P}=0.6$  are considered in this paper.
\begin{table}[t!]
 \centering\vspace{3mm}
\caption{Number $ N $ of dipoles for  particles with  porosity $\mathcal{P} $.}\vspace{-3mm}
\begin{tabular}{|c|ccccc|}\hline
 \large{$\mathcal{P}$}&$\mathbf{0.1}$&$\mathbf{0.2}$&$\mathbf{0.3}$&$\mathbf{0.4}$&$\mathbf{0.5}$\\\hline
$N$&467849&415866&363882&311899&259916\\\hline
\end{tabular}
  \label{sovieleNs}
\end{table}
%#########################################################################################################################################
%#########################################################################################################################################
%#####################################         2.4         ############################################################################### 
%#########################################################################################################################################
%#########################################################################################################################################
%#########################################################################################################################################
\subsection{$\beta$-ratio and blowout size}
\label{Motivation1}
We imagine a spherical grain in the orbit of a central star that is attracted by stellar gravitation and repelled by the radiation force. With decreasing radius, the radiation  force becomes increasingly important. As a consequence, there is a minimum particle radius $a_{\mathrm{BO}}$, called blowout size, at which the particle is blown out of the system and which can be calculated as a function of the  grains' optical properties, the stellar luminosity $L_*$, and stellar mass $M_*$ (\citealt{BurnsLamy}, \citealt{Artymowicz}). By modeling the density and grain size distribution of debris disks, a minimum grain radius is derived, which is often significantly above (a factor of $\sim$2 up to one magnitude in some cases) the value calculated under the assumption of compact grains (e.g. HD 104860, HD 8907, HD 377, HD 107146, HD 61005, HD 191089; \citealt{Roccatagliata2009}, \citealt{Ertel2011}). Despite model degeneracies, particularly between minimum grain sizes and inner disk radii, there have been several attempts to explain this discrepancy. A more detailed grain size distribution, derived in collisional evolution studies of debris disks and resulting in an overdensity of small particles (\citealt{Thebault2003}, \citealt{Krivov2006, Krivov2008}), can explain a factor of two. Grain segregation (\citealt{Roccatagliata2009}) or the occurrence of ice (\citealt{Greenberg1990}) affect the value of the blowout size as well (\citealt{Augereau2001}). However, another possible explanation for this problem can be the nature of the porous particles, which requires more investigations, since porosity influences the scattering and absorption properties of the particles (\citealt{Voshchinnikov1999}, \citealt{Vosh1, Vosh2}).
\begin{table}[b!]
	\centering
\vspace{3mm}
\caption{The physical properties of  main sequence stars (\citealt{BinneyMerrifield}) and respective stellar constant $C_{*}$ (see text for details).}
\vspace{-3mm}	
	\begin{tabular}{clllll}\hline\hline\vspace{-0.3cm}\\
\vspace{0.0cm}Spectral type&$\hspace{-0.15cm}\unit[T_{*}]{[K]}$&$\hspace{-0.3cm}M_{*}\hspace{0.05cm}[M_{\odot}]$&$\hspace{-0.3cm}R_{*}\hspace{0.05cm}[R_{\odot}]$	&$\hspace{-0.2cm}C_{*}\hspace{0.05cm}[C_{\odot}]$&$\hspace{-0.1cm}L_{*}\hspace{0.05cm}[L_{\odot}]$ \\\hline
A0	     &9520&$2.9$&$2.4$&$1.99$&$42.5$\\
A5	     &8200&$2.0$&$1.7$&$1.45$&$\enspace11.7$\\
F0	     &7200&$1.6$&$1.5$&$1.41$&$\enspace5.4$\\
F5	     &6440&$1.3$&$1.3$&$1.3$&$\enspace2.6$\\
G0	     &6030&$1.0$5&$1.1$&$1.15$&$\enspace1.4$\\
G5	     &5770&$0.9$2&$0.92$&$0.92$&$\enspace0.84$\\
K0	     &5250&$0.79$&$0.85$&$0.91$&$\enspace0.49$\\
K5	     &4350&$0.67$&$0.72$&$0.77$&$\enspace0.17$\\\hline
\end{tabular}
				 \label{CStern}
\end{table}
%#########################################################################################################################################
%#########################################################################################################################################
\subsubsection{Expression for the $\beta$-ratio}
We now derive an expression for the $\beta$-ratio, the ratio of radiation ($F_{\mathrm{rad}}$) to gravitational force ($F_{\mathrm{gra}}$), 
\begin{align}
\beta:=\frac{F_{\mathrm{rad}}}{F_{\mathrm{gra}}}. \label{GlBetaallg}
\end{align}
Under the assumption of an optically thin disk, the resulting radiation force is given by
\begin{equation} 
F_{\mathrm{rad}}=\frac{\pi^2 a^2}{c}\frac{R_{*}^2}{r^2}\int Q_{\mathrm{pr}}(\lambda)\,B_{\lambda}(T_*) \,\mathrm d\lambda,\label{GlFrad}
\end{equation}
(\citealt{Koehler02, Koehler04}) where  the star is considered as a black body with  effective temperature $T_*$,  radius $R_{*}$ and  mass $M_{*}$. The quantity $B_{\lambda}(T_*)$ is the Planck function, $c$ is the speed of light, and $Q_{\mathrm{pr}}(\lambda)$ is the wavelength-dependent efficiency factor of the radiation pressure,
\begin{equation}
Q_{\mathrm{pr}}=Q_{\mathrm{abs}}+Q_{\mathrm{sca}}(1-g),
\end{equation}
which is a function of the absorption ($Q_{\mathrm{abs}}=C_{\mathrm{abs}}/(\pi a^2)$) and scattering cross section ($Q_{\mathrm{sca}}=C_{\mathrm{sca}}/(\pi a^2)$), and the  asymmetry factor  $g$. For a given grain mass $m_{\mathrm{g}}= 4/3\pi a^3(1-\mathcal{P})\rho$ the gravitational force on the particle amounts to
\begin{equation} 
F_{\mathrm{gra}}=\gamma\, \frac{M_{*}m_{\mathrm{g}}}{r^2}=\gamma\, \frac{4\,\pi\,M_{*}\, a^3\,\rho}{3\,r^2} (1-\mathcal{P}), \label{GlFgra}
\end{equation}
where $\gamma$ is the gravitational constant. Using the Eqs.  \ref{GlBetaallg}, \ref{GlFrad} and \ref{GlFgra} delivers
\begin{equation}
\beta=\,C_{*}  \frac{1}{a} \frac{1}{(1-\mathcal{P})} \frac{1}{\rho} \int Q_{\mathrm{pr}}(\lambda)\, B_{\lambda}(T_*) \,\mathrm d\lambda,
\label{betaGleichung}
\end{equation}
with
\begin{equation}
C_{*}:=\unit[117.76]{\frac{kg\,s^3}{m^4}}\,\frac{R_{*}^{2}}{M_{*}}.\label{Nikolaus}
\end{equation}
For the Sun one thus derives $C_{*}=C_{\odot}=2.865 \cdot \unit[10^{-11}]{s^{3}\,m^{-2}}$. The quantity $C_{*}$ is a function of stellar parameters alone, while the other factors in front of the integral in Eq. \ref{betaGleichung} represent the dust properties. The integral reflects the interaction between dust and star. In Table \ref{CStern}, the factor  $C_{*}$ is presented  for main sequence stars of spectral type A to K. The stellar parameters are taken from \citet{BinneyMerrifield}, as published in \citet{SchmidtKaler}.

The particle is ejected from the system, if $\beta (a)>1/2$. A particle that initially moves on a circular orbit is compelled  on an elliptic orbit for $\beta <0.5 $. The trajectory is parabolic (hyperbolic) for $ \beta = 0.5\hspace{0.2cm} (> 0.5 )$, and the particle leaves the system (\citealt{Krivov2006}).

We would like to remind the reader that in the literature (see \citealt{BurnsLamy}, \citealt{Artymowicz}, \citealt{Kruegel}, \citealt{Hahn}) an approximation equation is usually  given that neglects the wavelength dependence of $Q_{\mathrm{pr}}$. In the limit of the geometrical optics, $Q_{\mathrm{pr}}$ is equal to 1, resulting in
\begin{align} 
\beta= 0.577\left(\frac{\rho}{\mathrm{g}\, \mathrm{cm}^{-3}}\right) ^{-1} \left(\frac{a}{\mu \mathrm{m}}\right)^{-1} \left(\frac{L_{*}}{L_{\odot}}\right) \left(\frac{M_{*}}{M_{\odot}}\right)^{-1}. \label{betakurzundeinfach}
\end{align}
Equation  \ref{betakurzundeinfach} is easy to handle and  therefore suitable for a  rough estimation of the blowout size. However, for more precise studies (as required in our investigations)  the wavelength dependence of $Q_{\mathrm{pr}}$ expressed in Eq. \ref{betaGleichung} must be considered.
%#########################################################################################################################################
%#########################################################################################################################################
\subsubsection{Previous studies}
The influence of porosity on the radiation pressure has already been considered  in previous studies. 
\cite{Mukai92}, \cite{Kimura1997}, and \cite{Koehler2007} have examined porous aggregates with fractal structures, ballistic particle-cluster aggregates (BPCA), and ballistic cluster-cluster aggregates (BCCA), which are described in terms of the fractal dimension (e.g. \citealt{Witten1986}).
These aggregates are two extreme cases of relatively compact and highly porous aggregates, whereas we apply a grain model that allows us to increase the porosity step by step, permitting an analysis of the physical dust properties with changing porosity.
\cite{Mukai92} investigated the $\beta$-ratio based on Maxwell Garnett effective medium theory (\citealt{Garnett1904}), and \cite{Kimura1997} used the Bruggeman rule (\citealt{Bruggeman1935}). \cite{Hage1990} assessed the accuracy of $Q_{\mathrm{abs}}$ when determined with Maxwell Garnett, while \cite{Bazell1990} report a discrepancy at longer wavelengths. \cite{Voshchinnikov1999} found that effective medium theories are  considerably worse for calculating scattering cross sections and the asymmetry factor $g$ than for extinction cross sections.

In \cite{Koehler06,Koehler2007} the DDA is applied to simulate light scattering by irregular aggregates. They calculated the optical properties of larger grains only
 for one wavelength ($\lambda=\unit[0.6]{\mu m}$) and thus neglect the wavelength dependence of $Q_{\mathrm{pr}}$. In our study the number of dipoles is much larger, which allows us to calculate the optical properties for a wide wavelength range and for grains up to $a=\unit[4]{\mu m}$.  We then use the obtained data to investigate not only the $\beta$-ratio, but also the blowout size for various porosities, stellar temperatures, and number of dipoles, and simultaneously calculate the dust temperature of porous grains, especially the one of the blowout grains. This gives a comprehensive overview of porous grains in debris disks.
%#########################################################################################################################################
%#########################################################################################################################################
%#####################################         3           ############################################################################### 
%#########################################################################################################################################
%#########################################################################################################################################
%#########################################################################################################################################

\section{Results for the efficiency factor of absorption}
\label{firstres}
\begin{figure}[t!]
	\centering
\includegraphics[width=1.0\linewidth]{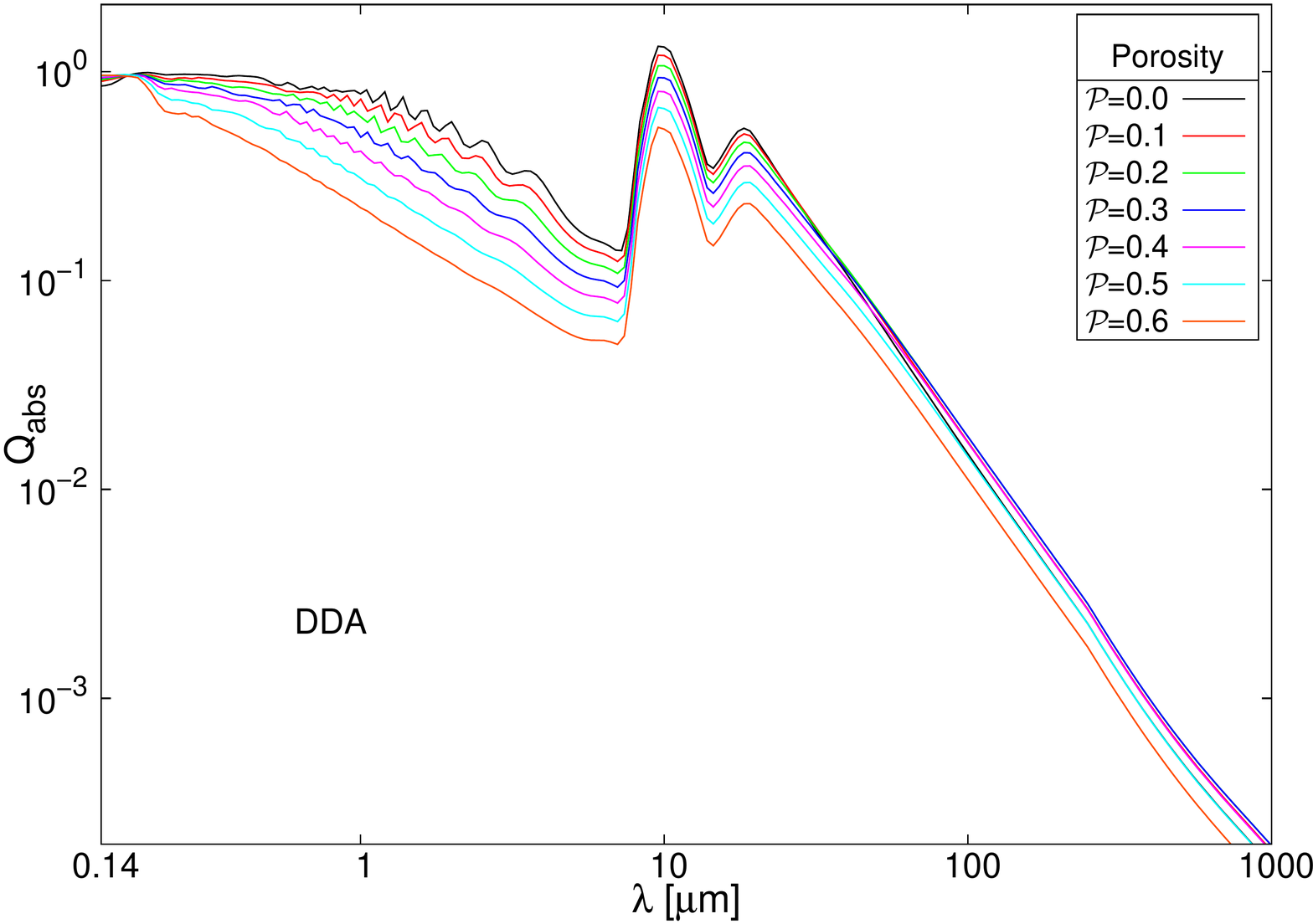}
\includegraphics[width=1.0\linewidth]{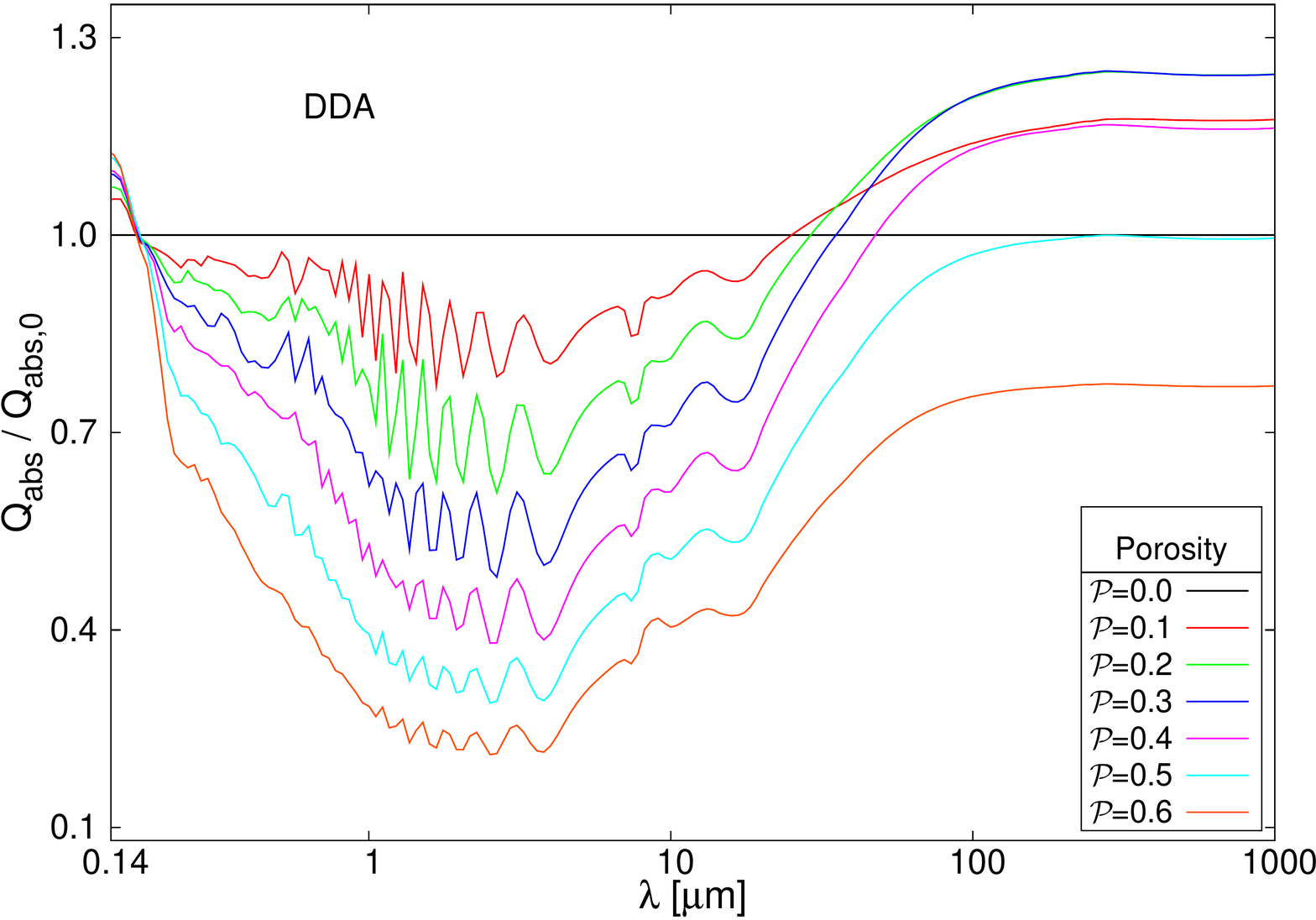}
	\caption{Wavelength-dependent absorption cross section $Q_{\textup{abs}}$ as a function of porosity $\mathcal{P}$. The particle radius is 
$a=\unit[1.0]{\mu m}$.  Note the two characteristic silicate peaks  at $\lambda\sim\unit[10]{\mu m}$  and $\unit[18]{\mu m}$. In the bottom image the results are normalized to the values of compact spheres, $Q_{\textup{abs},0}$.}
	\label{qabsverpor}
\end{figure}

The $\beta$-ratio of a dust grain is a function of the optical properties. The simulation software DDSCAT delivers the quantities $Q_{\mathrm{abs}}$, $Q_{\mathrm{sca}}$, and $g$ for each wavelength and radius, therefore we investigate the absorption cross sections  for porous  particles with radius  $a= \unit[1]{\mu m}$ (upper image of  Fig. \ref{qabsverpor}). The porosity varies in steps of $0.1$ from $\mathcal{P} = 0.0 $ to $0.6$. To emphasize the dependence on porosity, the curves are normalized to the values of compact spheres ($Q_{\textup{abs},0}$) in the bottom image of Fig. \ref{qabsverpor}.  The relative differences of $Q_{\textup{abs}}$ increase with porosity for all wavelengths lower than $\unit[25]{\mu m}$. $Q_{\textup{abs}}$ drops with increasing porosity for wavelengths from $0.2$ to $\unit[25]{\mu m}$ and increases for $\lambda<\unit[0.2]{\mu m}$. For $\lambda>\unit[50]{\mu m}$, $Q_{\textup{abs}}$ increases with porosity up to $\mathcal{P}=0.3$  and  then decreases again. The values for the porosity $\mathcal{P} = 0.6 $ are $ 0.21$ to $1.13$ times larger than the ones for compact spheres. Around $\lambda\sim\unit[1]{\mu m}$, the curves have a wavy pattern, which diminishes with increasing porosity.

Furthermore, we calculated the absorption cross sections of the same grain with an effective medium theory (Fig. \ref{qabsverporEMT}), using the Bruggeman rule (\citealt{Bruggeman1935}). Besides the main similarity of both figures, the main differences are found for $\lambda<\unit[0.2]{\mu m}$ and  $\lambda>\unit[20]{\mu m}$. The values for the porosity $\mathcal{P} = 0.6 $ are $0.25$ to $1.2$ times larger than the ones for compact spheres. In addition, the wavy pattern around $\lambda\sim\unit[1]{\mu m}$ is more pronounced and can be detected also for $\mathcal{P} = 0.6$.

%#########################################################################################################################################
%#########################################################################################################################################
%#########################################################################################################################################
%#########################################################################################################################################
%#########################################################################################################################################
%#########################################################################################################################################
%###########################################            4        #########################################################################
%#########################################################################################################################################
%#########################################################################################################################################
%#########################################################################################################################################
\section{Blowout size for different parameter}
\label{Kapitel4}
We determine the blowout size as a function of the porosity and the stellar luminosity.
%__________________________________________________________________________________________________________________________________________________________________________
%__________________________________________________________________________________________________________________________________________________________________________
%_______________________________________    4.1  __________________________________________________________________________________________________________________________
%__________________________________________________________________________________________________________________________________________________________________________

\subsection{Blowout size for porous particles in the solar system}
\label{Abschnitt36}
\begin{figure}[t!]
 \centering
\includegraphics[width=1.0\linewidth]{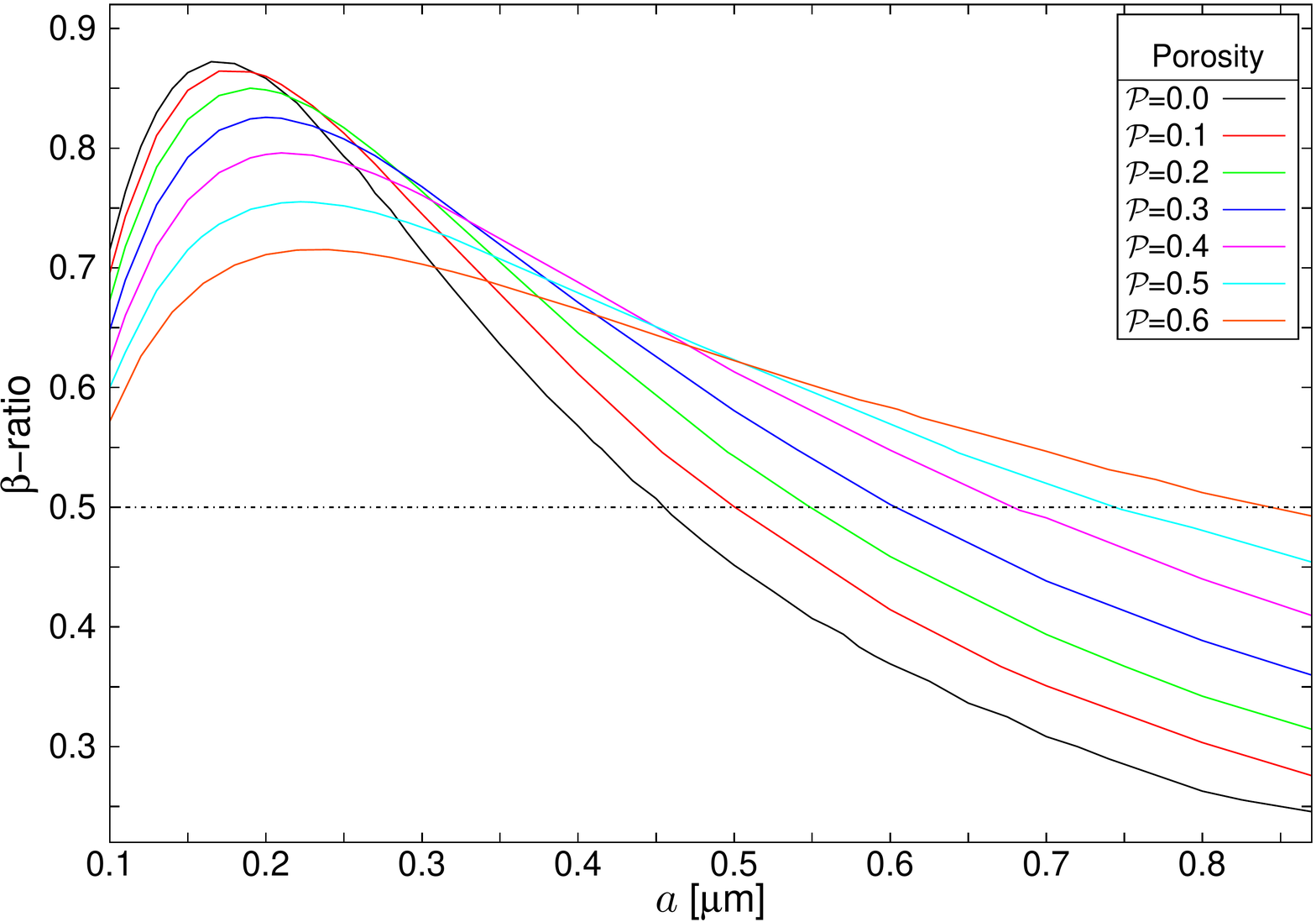}
\caption{The $\beta$-ratio as a function of the grain radius $a$ and porosity $\mathcal{P}$ for the solar system. The intersections of the different distributions  with the 0.5 limit mark  the blowout size $a_{\mathrm{BO}}$.}
\label{BetaSonne}
\end{figure}

\begin{figure}[t!]
	\centering
\includegraphics[width=1.0\linewidth]{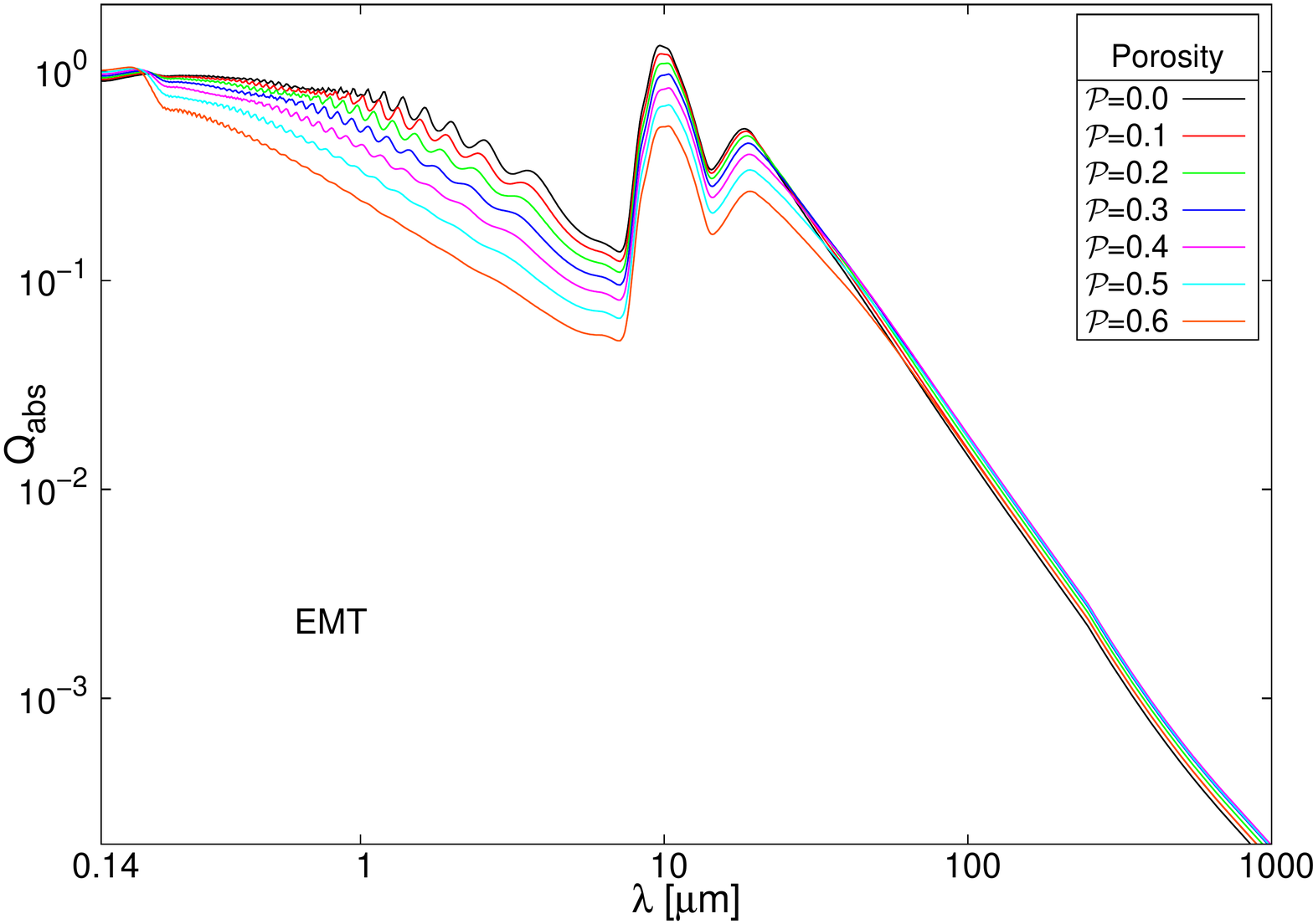}
\includegraphics[width=1.0\linewidth]{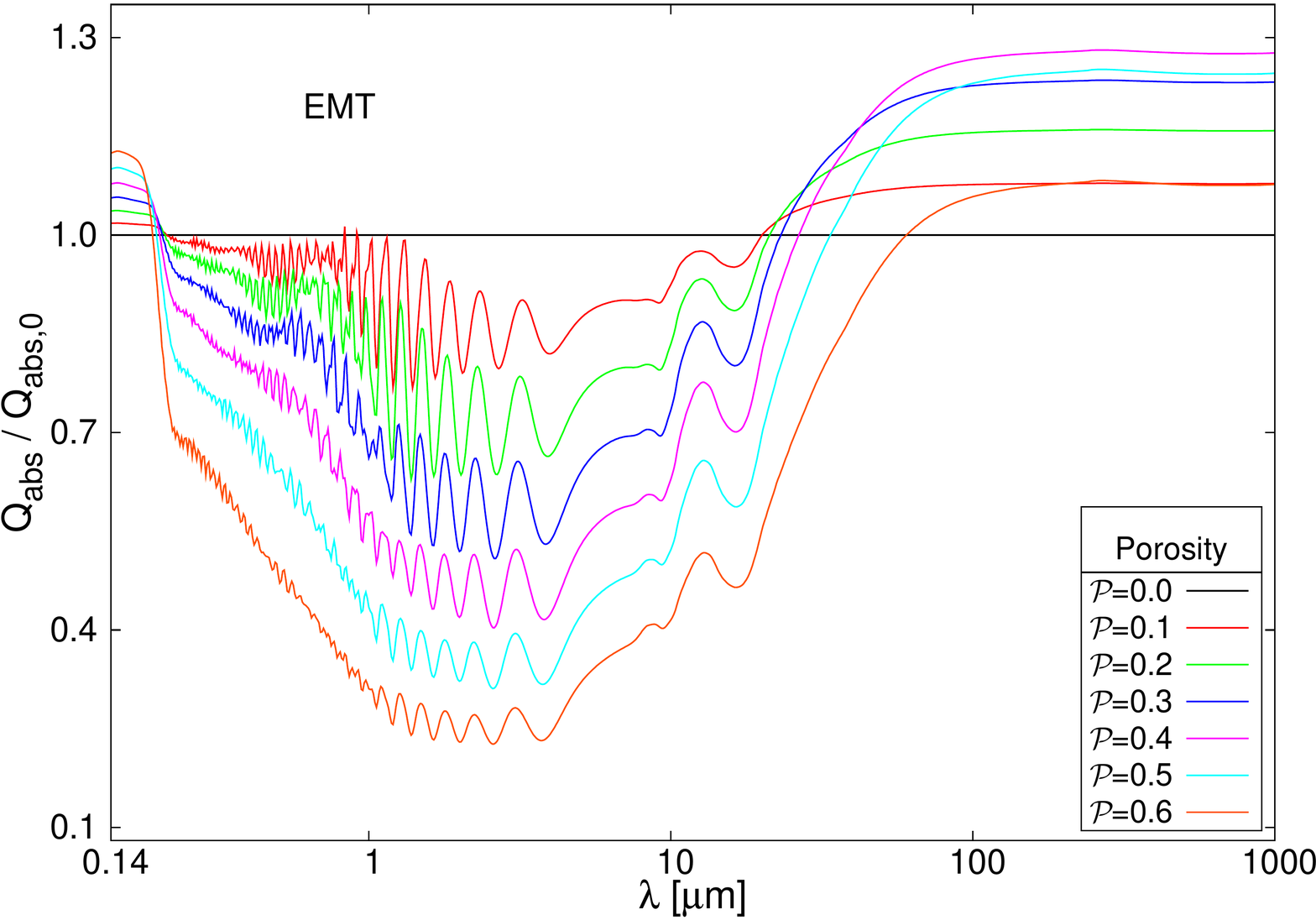}
	\caption{Same as Fig. \ref{qabsverpor}, but derived with effective medium theory (EMT, Bruggeman rule).\newline~\newline~\newline~}
	\label{qabsverporEMT}
\end{figure}

At first we investigate the impact of grain porosity on the  blowout size of the solar system. For this purpose, the $\beta$-distributions of compact spheres and porous particles are determined. The effective temperature of the Sun is $T_*=\unit[5777]{K}$, and the porosity  varies in steps of $0.1$ from $\mathcal{P} = 0.0 $ to $0.6$. The distributions are presented in Fig. \ref{BetaSonne}, and the corresponding blowout sizes are listed in Table \ref{BO0}.

\begin{table}[b!]
\renewcommand{\arraystretch}{1.15}
	\centering
\vspace{3mm}	\caption{Blowout size as a function of grain porosity $\mathcal{P}$ in the solar system.}\vspace{-3mm}
		\begin{tabular}{c|c|c|c|c|c|c}
$\mathcal{P}$              &$0.0$&$0.1$&$0.2$&$0.3$&$0.4$&$0.5$\\\hline
$a_{\mathrm{BO}}\,[\mu$m$]$&$0.455$&$0.500$&$0.549$&$0.603$&$0.678$&$0.744$
\end{tabular}
		 \label{BO0}
\end{table}
            \begin{figure}[h!]
 \centering
\includegraphics[width=1.0\linewidth]{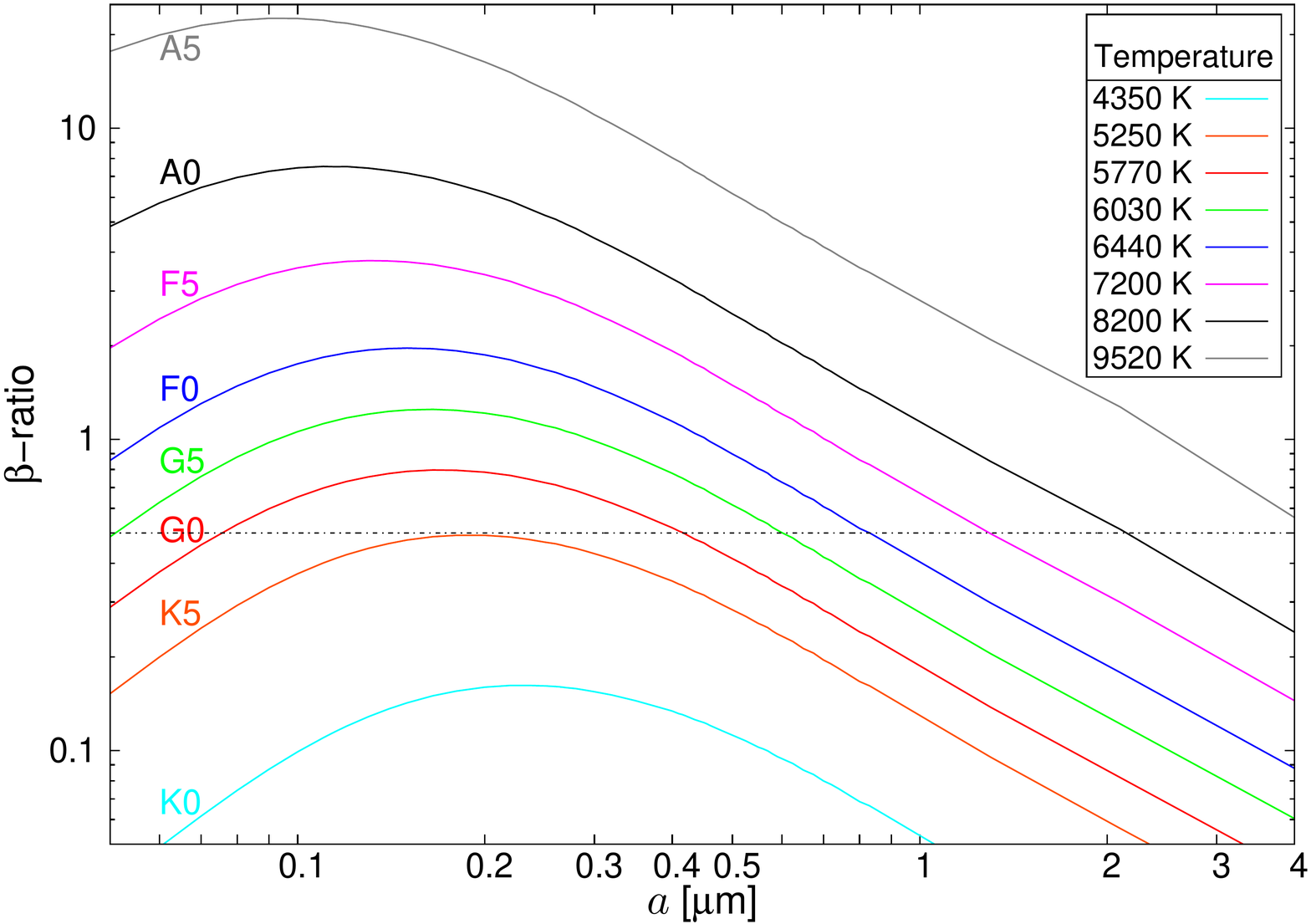}
\caption{The $\beta$-ratio as a function of the grain radius $a$ for stars  of various spectral type (different stellar temperature). The particles are compact spheres.}
\label{BetaTemp}
\end{figure}

\begin{enumerate}
\item The $\beta$-value strongly increases with radius, reaching a maximum, and then starts falling off. With increasing porosity, the maximum of the $\beta$-distribution decreases and shifts towards larger radii. Thus, the maximum is the largest for compact spheres. Furthermore, the slope at the larger radii becomes flatter.
 \item The blowout size of compact spheres is $a_{\mathrm{BO}}=\unit[0.46]{\mu m}$. For comparison, the  value  $\unit[0.33]{\mu m}$ is obtained with the approximation Eq. \ref{betakurzundeinfach}. In \citet{Mukai92}  a blowout size between $\unit[0.3]{\mu m}$ and $\unit[0.35]{\mu m}$ can be estimated, and in \citet{Koehler02} between $\unit[0.45]{\mu m}$ and $\unit[0.5]{\mu m}$ (see paper, Figs. 4b and 1, resp.).
\item The higher the porosity, the larger the blowout size. For $ \mathcal{P} = 0.6 $ the blowout size is $ a_{\mathrm{BO}} = \unit[0.84]{\mu m}$, $1.8$ times larger than for $\mathcal{P} =0.0$. This factor is comparable to the factor between the minimum grain size obtained with disk modeling and the value calculated under the assumption of compact spheres.
\end{enumerate}

%__________________________________________________________________________________________________________________________________________________________________________
%__________________________________________________________________________________________________________________________________________________________________________
%_________________________________________  4.2  __________________________________________________________________________________________________________________________
%__________________________________________________________________________________________________________________________________________________________________________

\subsection{Blowout size of stars of various spectral type}
\label{Abschnitt37}
The influence of different stellar luminosities on the blowout size is investigated. For this purpose, we consider compact spherical grains with  particle radii up to $\unit[4]{\mu m}$, since higher luminosities increase the radiation force and thus the blowout size. The values ​​for $T_*$ and $C_{*}$ for the eight different spectral types from K0 to A5 are summarized in Table \ref{CStern}. The  resulting $\beta$-distributions are shown in Fig. \ref{BetaTemp}.
\begin{enumerate}
 \item The higher the luminosity, the higher the $\beta$-values, whereby the blowout size increases. In addition, the maximum of the\newpage~\newpage~\newpage $\beta$-distribution shifts to smaller particle radii for higher luminosities.
 \item For some luminosities one finds $ \beta <0.5 $ for all particle radii, so there no blowout size exists in these systems. In particular, only stars
with temperatures $T_*>\unit[5250]{K}$ have a blowout size. There is also no  blowout size at lower temperatures for porous grains, since the maximum  of the $\beta$-distribution is the greatest for compact spheres, and it decreases with increasing porosity (Sect. \ref{Abschnitt36}).
 \item The blowout size  at $T_*=\unit[9520]{K}$ is larger than $\unit[4]{\mu m}$, and even higher values​​ are expected for porous  particles. Calculations of the optical properties of such large grains exceed our computational capabilities, therefore stars with $T_* \ge\unit[9520]{K}$ are not investigated further.
 \item Several temperature curves intersect the $0.5$ limit  in the examined particle radii interval even twice, whereby two values for the blowout size exist (e.g. the $T_*=\unit[5770]{K}$ curve). Further forces are acting on  very small particles, such as the Poynting-Robertson effect and the Lorentz force, which are negligible for larger grains, but have to be taken into account for the balance of forces on very small grains (\citealt{Grun2001}).
\end{enumerate}

%__________________________________________________________________________________________________________________________________________________________________________
%__________________________________________________________________________________________________________________________________________________________________________
%_____________________________________________________  4.3    ____________________________________________________________________________________________________________
%__________________________________________________________________________________________________________________________________________________________________________

\subsection{Parameter space}
\label{Abschnitt38}
\begin{figure}[t!]
 \centering
\includegraphics[width=1.0\linewidth]{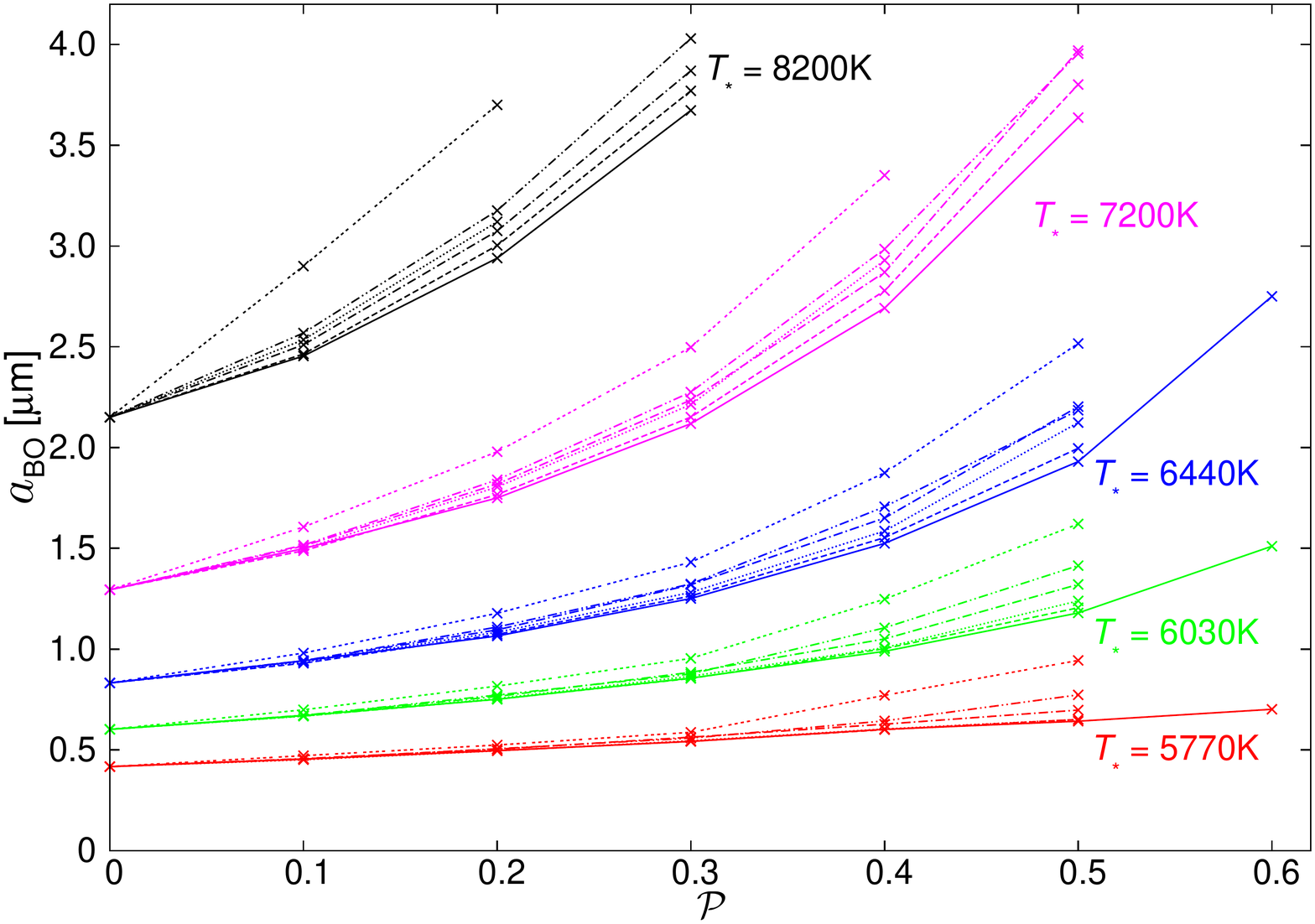}
\caption{Blowout size $a_{\textrm{BO}}$ as a function of porosity $\mathcal{P}$ and stellar temperature $T_*$ for various number of dipoles. Different  colors represent different  temperatures. The values for the special particle presented in Sect. \ref{genera} are illustrated as the solid line.}
\label{sechsL}
\end{figure}

As shown in the previous two sections, the blowout size depends on both the grain  porosity and the stellar temperature. We now determine the blowout size for each parameter configuration. The computational effort is large, especially  particles with radii $ a> \unit[2]{\mu m} $ increase the calculation time dramatically. We generate and evaluate the data for the following parameter space:
\begin{itemize}
 \item seven porosities: $\mathcal{P}=0.0$, $0.1$, $0.2$, $0.3$, $0.4$, $0.5$, and $0.6$,
 \item five stellar temperatures: $T_*=\unit[5770]{K}$, $\unit[6030]{K}$, $\unit[6440]{K}$, $\unit[7200]{K}$, and $\unit[8200]{K}$,
 \item six number of dipoles, from $54,432$ to $519,832$ dipoles.
\end{itemize}
We vary the total number of dipoles and thus the size of the inner voids to investigate the influence of different particle structures. 
The calculated blowout sizes of the special particle presented in Sect. \ref{genera} are illustrated in Fig. \ref{sechsL} and listed in Table \ref{RESULT}. Missing table entries imply a blowout size larger than $\unit[4]{\mu m}$ and are thus not computable within the parameter settings.

Figure \ref{sechsL} again reveals the blowout size increasing  both with grain porosity and stellar temperature. For  $ T_* =\unit[5770]{K}$ the blowout size of porous grains ($\mathcal{P} = 0.5 $) is larger by a factor of $1.5$  if compared to compact spheres. The higher the temperature, the stronger the effect, so that  the factor amounts to $2.8$ for $T_* = \unit[7200]{K}$. These factors  are close to matching  the existing discrepancy mentioned in Sect. \ref{Motivation1}. The variation in the number of dipoles causes a dispersion of the blowout sizes. This dispersion increases also both with $\mathcal{P}$ and $T_*$. In most cases, $a_{\mathrm{BO}}$ rises with decreasing number of dipoles. However, there are exceptions. The largest deviations from the other calculated blowout sizes occur for the lowest number of dipoles, the coarsest particles, whose $a_{\mathrm{BO}}$ are significantly larger than those from the other ones (triple dotted lines).

Figure \ref{sechsL} can be used to estimate  the blowout size for a given grain porosity $ \mathcal{P}  $ and stellar temperature $T_*$. The different blowout sizes  at constant porosity and temperature, caused by the varying number of dipoles, set the range of applicability in which the blowout size  can be found.
\begin{table}[t!]
\centering
\vspace{3mm}\caption{Blowout size in $\mu$m as a function of grain porosity $\mathcal{P}$ and  stellar temperature $T_*$.}\vspace{-3mm}
\begin{tabular}{|c|ccccc|}\hline
&\multicolumn{5}{c|}{$\unit[T_*]{[K]}$}\\
$\mathcal{P}$&5770&6030&6440&7200&8200\\\hline
\vspace{-3mm}&&&&&\\
 $\mathbf{0.0}$&$0.42$&$0.60$&$0.83$&$1.29$&$2.15$\\ 
$\mathbf{0.1}$&$0.45$&$0.67$&$0.94$&$1.50$&$2.45$\\
$\mathbf{0.2}$&$0.50$&$0.75$&$1.07$&$1.75$&$2.94$\\
$\mathbf{0.3}$&$0.54$&$0.86$&$1.25$&$2.12$&$3.67$\\
$\mathbf{0.4}$&$0.60$&$0.99$&$1.52$&$2.69$&\\
$\mathbf{0.5}$&$0.64$&$1.18$&$1.93$&$3.64$&\\
$\mathbf{0.6}$&$0.70$&$1.51$&$2.75$&&\\\hline
\end{tabular}
			 \label{RESULT}
\end{table}
%__________________________________________________________________________________________________________________________________________________________________________
%__________________________________________________________________________________________________________________________________________________________________________
%____________________________________________   4.4   _____________________________________________________________________________________________________________________
%__________________________________________________________________________________________________________________________________________________________________________
\subsection{Fitting model}
\label{Abschnittfit}
In this section  an approximation function  $ f \, (\mathcal{P}, T_* $) for the  blowout size is  derived. We find that the results of the special particle presented in Sect. \ref{genera} can be analytically approximated very well using the following equation:
\begin{align}
a_{\mathrm{BO}}\left(\mathcal{P},\,T_*\right)=(k_1 T_*+&k_2) \cdot \mathcal{P}^{\alpha}\,+\,\left(k_3 T_*+k_4\right), \label{GlFit}\\[0.3cm]
k_1=\unit[5.07\cdot 10^{-3}]{\frac{\mu m}{K}},&\quad k_2=\unit[-28.85]{\mu m},\nonumber\\[2mm]
k_3=\unit[7.01\cdot 10^{-4}]{\frac{\mu m}{K}},&\quad k_4=\enspace\unit[-3.65]{\mu m},\nonumber\\[2mm]
\alpha=1.781-&\frac{61.72\,\mathrm{K}}{T_*\,-\,5670\, \mathrm{K}}.\nonumber 
\end{align}
It contains seven coefficients (three for $\alpha$, and the $k_i$,  $i \in \mathbb{N}_{\le 4}$).

\begin{figure}[t!]
 \centering
\includegraphics[width=1.0\linewidth]{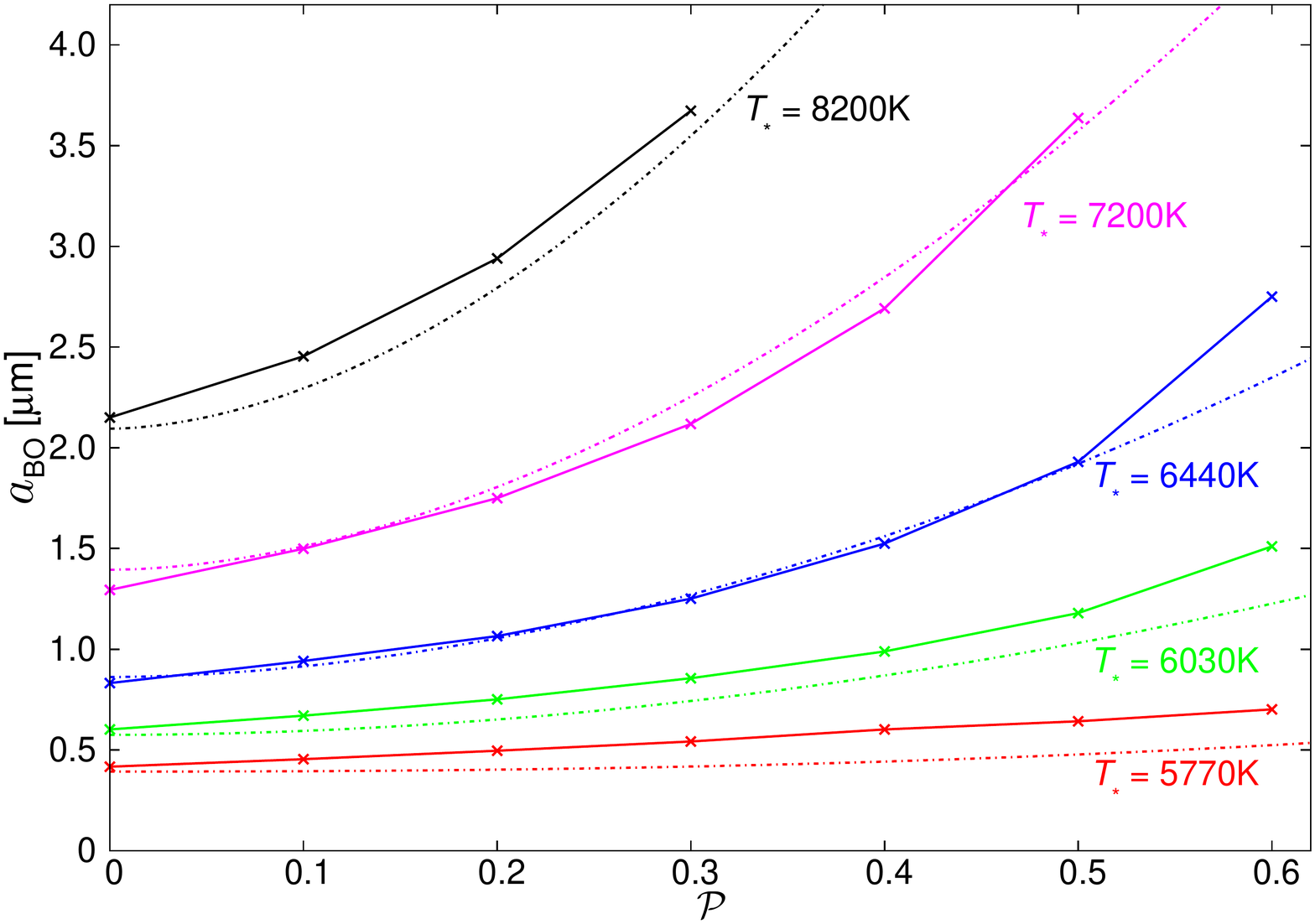}
\caption{Calculated blowout sizes (solid line) and derived approximation function (dotted lines). Using the equation, it is possible to estimate $a_{\mathrm{BO}} $ as a function of grain porosity and stellar temperature.
}
\label{Fitmodell}
\end{figure}

The fit is  presented in Fig. \ref{Fitmodell}. The relative deviation  between fit and measured values is for the highest temperature lower than $8\,\%$, so Eq. \ref{GlFit} is a good way to estimate the size of the blowout. An online tool for calculating the blowout size based on this approximation function is available at
\\ \textit{http://www1.astrophysik.uni-kiel.de/blowout/}.

%__________________________________________________________________________________________________________________________________________________________________________
%__________________________________________________________________________________________________________________________________________________________________________
%_____________________________________   4.5   ____________________________________________________________________________________________________________________________
%__________________________________________________________________________________________________________________________________________________________________________
\subsection{$\beta$ influenced by other effects}
\label{Abschnitt39}
We discuss other possible effects which might influence the $\beta$-ratio.

\subsubsection{Particle modeling}
The particles  used in this paper have a spherical shape, into which holes are formed. Besides prolate and oblate shapes, real particles can have a much more complex non symmetric structure. However, the vast variety of particle shapes cannot be examined. The spherical shape with inclusions was chosen  as representative of all porous and arbitrary shaped particles.

It should be noted that large particles are generally more porous than smaller ones, because they can be seen as agglomerates of smaller subunits.
\citet{LamyGrun} modeled  the density of small grains by $\rho=2.2-1.4\,(a^2+a)$, where $a$ is in $\mu$m and $\rho$ in $\textup{g}\, \textup{cm}^{-3}$. With increasing particle radius this  density law leads to higher $ \beta $-values.

There is no unique definition of the optical properties and bulk densities of what is considered as an ``astronomical silicate``. As the $\beta $-ratio is inverse to the density, the choice of $\rho$ has a strong impact on the blowout size. It is  also conceivable that individual particles consist of several materials.

The mentioned  points   do not lead to errors in the calculations. Instead, they show the possible complexity  and variety of  particles in the disk, hence the difficulty in specifying the blowout size of a  stellar system, since a separate  $\beta $-value can be allocated to each particle and  even to each particle orientation.

\subsubsection{Central star}
The stellar component is considered as a black body with the effective temperature $T_*$. In contrast to this, \citet{Lamy74b} and  \citet{Artymowicz} motivated to use an observed  or a modeled photospheric spectrum (e.g. Kurucz model). We did not consider this in our parameter space study because fundamentals are first investigated. In the treatment of a real object, the shape of the spectrum should be used. Furthermore, $C_{*}$ is calculated for parameters of a main-sequence star model. Real stars can deviate from these values​​, so real values should be used for stellar mass and radius if available.
\subsubsection{Ejection limit}
The ejection limit $\beta_{\mathrm{ejection}}=0.5$ only applies to particles  that have been produced by planetesimals that move in circular orbits, i.e. for which the orbit eccentricity is $e  = 0.0$. The general $\beta$ threshold depends on the position $\Phi$ of the elliptical orbit, where the particle is set free, and is  $  \beta_{\Phi,\mathrm{ejection}}=\frac{1}{2}\left(\frac{1\,-\,e^2}{1\,+\,e\,\cos{(\Phi})}\right)$. Even  eccentricities as low as $ e = 0.1 $ can cause $ \beta$-values ​​from $ 0.45 $ to $0.55 $. Owing to the different  ejection conditions for each particle, the blowout size depends on the eccentricity. This influence  is  quite strong compared to the other effects discussed above. 
\subsubsection{Further  forces acting on dust grains in debris disks}
The derivation of Eq. \ref{betaGleichung} for  $\beta$ only considers gravitational and radiation forces. In addition, the Poynting-Robertson effect and stellar wind drags are acting  on the particles, and  in the case of charged particles also the Lorentz force. The magnitude of these forces depends strongly on the particle size. The gravitational force dominates for particles with radii $a>\unit[1]{\mu m}$. For particles with radii $\unit[0.05]{\mu m}$ or lower, the Lorentz force prevails. For intermediate particle radii, the radiation force is important   beside gravitation and Lorentz force. In \citet{Grun99}, the various forces  are listed and compared for different particle radii.

%#########################################################################################################################################
%#########################################################################################################################################
%###########################################            5        #########################################################################
%#########################################################################################################################################
%#########################################################################################################################################
%#########################################################################################################################################

\section{Dust temperature of fluffy grains}
\label{Staubtemp}
\begin{figure}[h!]
 \centering
\includegraphics[width=1.0\linewidth]{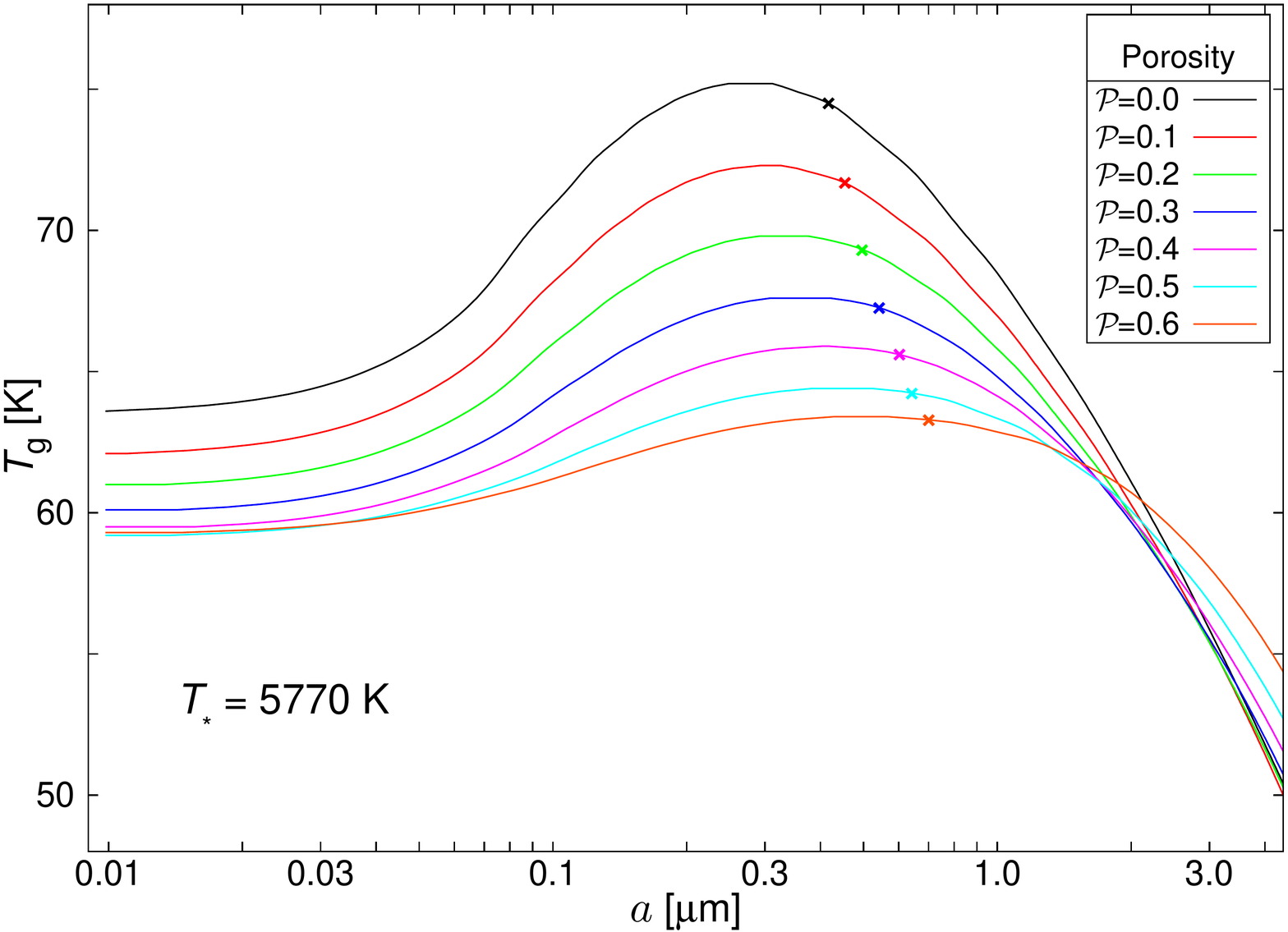}
\caption{Dust temperature $T_{\mathrm{g}}$ at the distance of $\unit[50]{AU}$ from a star with the effective temperature $T_*=\unit[5770]{K}$, depending on the grain radius $a$ and porosity $\mathcal{P}$. The blowout sizes are marked as crosses.}
\label{TempT=5770K}
\includegraphics[width=1.0\linewidth]{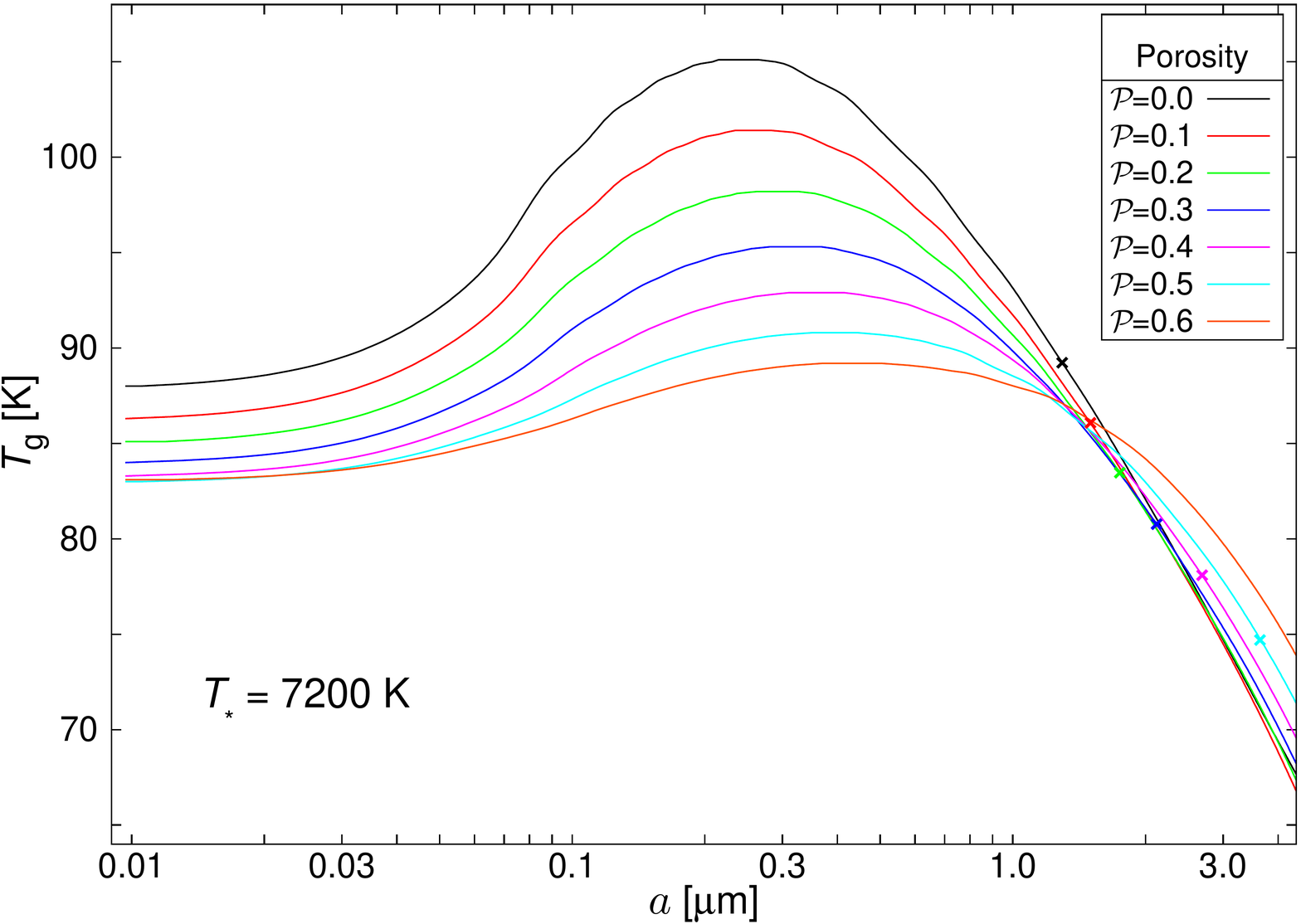}
\caption{Same as Fig . \ref{TempT=5770K}, only for the stellar temperature $T_{*}=\unit[7200]{K}$.}
\label{TempT=7200K}
\end{figure}

Finally, the influence of porosity on the dust equilibrium temperature is investigated. For this purpose the dust temperature $T_{\mathrm{g}}$ is determined if the grains are only heated by direct stellar radiation. When taking the energy conservation into account, the distance $r$ of the particle from the star is
\begin{equation}
r=\frac{R_*}{2}\sqrt{\frac{\int_0^\infty B_{\lambda} \left(T_*\right) Q_{\mathrm{abs}}\left(a,\lambda\right)  \,\mathrm d\lambda }{\int_0^\infty B_{\lambda}\left(T_{\mathrm{g}}\right) Q_{\mathrm{abs}}\left(a,\lambda\right)  \,\mathrm d\lambda }}\, ,\label{AbstanddT}
\end{equation}
(\citealt{WolfHillenbrand2003}). We calculated the dust temperature at a distance of $r=\unit[50]{AU}$ for stellar temperatures $T_*=\unit[5770]{K}$ and $T_*=\unit[7200]{K}$ and illustrate them in Figs. \ref{TempT=5770K} and \ref{TempT=7200K}, respectively.

\begin{figure}[h!]
 \centering
\includegraphics[width=1.0\linewidth]{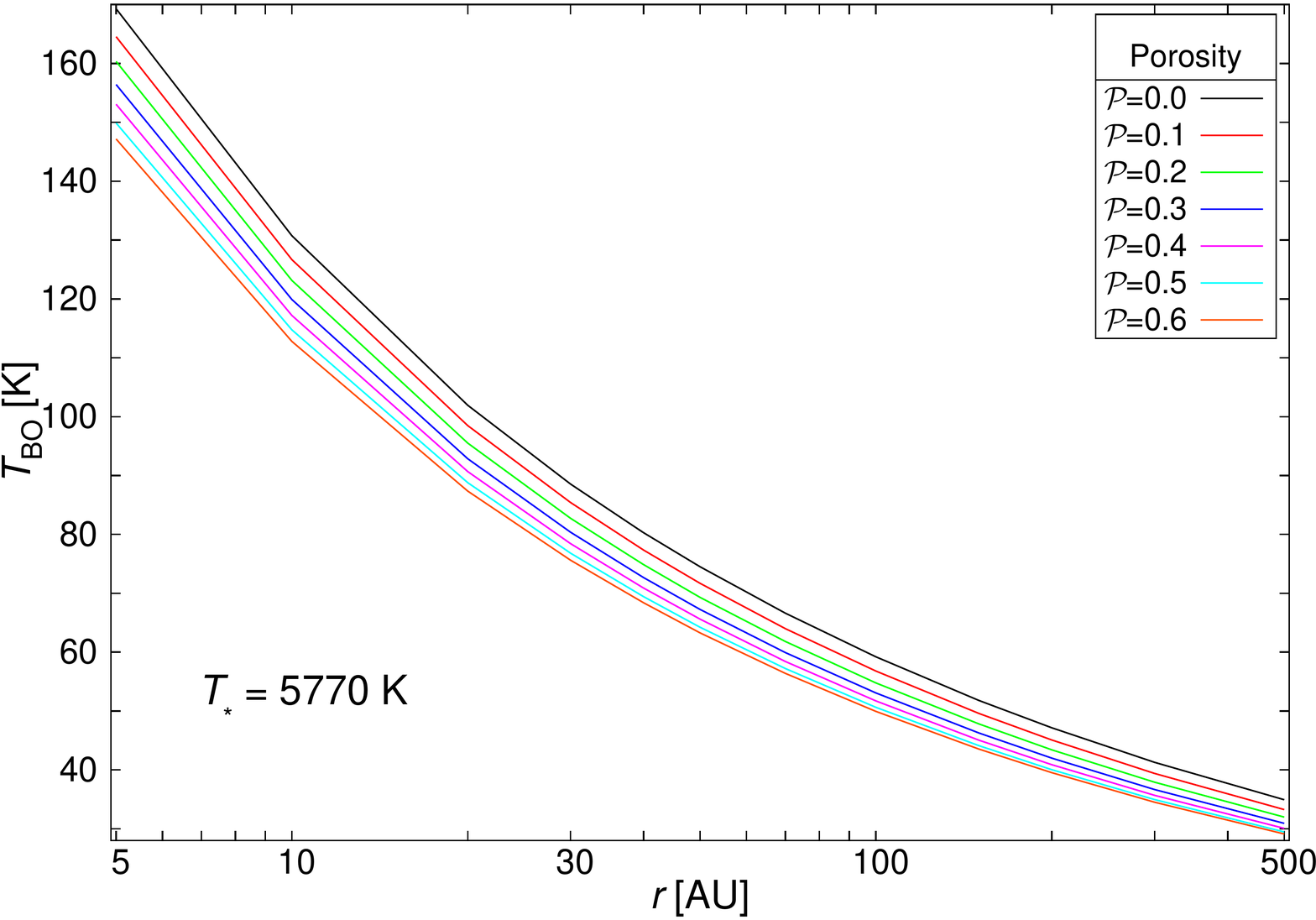}
\caption{Dust temperature of the blowout size grains $T_{\mathrm{BO}}$ as a function of porosity $\mathcal{P}$ and the distance $r$ from the central star. The stellar temperature is $T_*=\unit[5770]{K}$.}
\label{TaBO}
\end{figure}
\begin{figure}[h!]
\includegraphics[width=1.0\linewidth]{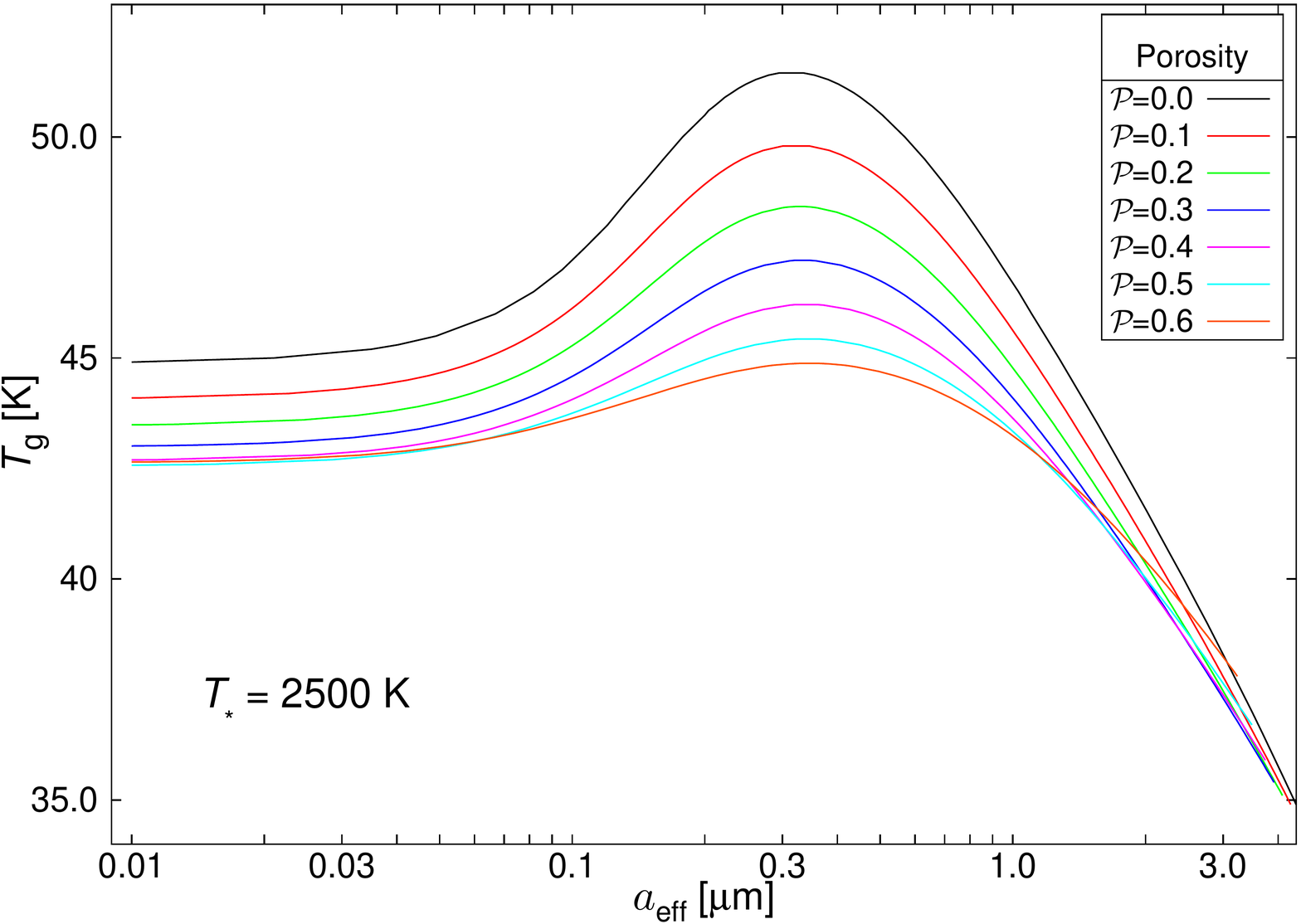}                       
\caption{Dust temperature $T_\textup{g}$ at the distance of $10^4$ stellar radii from a star with effective temperature $T_*=\unit[2500]{K}$, depending on the porosity and the effective radius  $a_{\mathrm{eff}}=(1-\mathcal{P})^{\nicefrac{1}{3}}\, a$. The dust material is a 1:1 mixture of astronomical silicate and amorphous carbon.}
\label{TempAC1}
\end{figure}
\begin{enumerate}
\item The temperature of the smallest particles  increases with the grain radius up to a maximum before it falls off steeply for larger radii. In addition, the maximum shifts with increasing porosity from $\unit[0.28]{\mu m}$ for $\mathcal{P}=0.0$ to $\unit[0.5]{\mu m}$ for $\mathcal{P}=0.6$ in the case of $T_*=\unit[5770]{K}$, and from $\unit[0.24]{\mu m}$ for $\mathcal{P}=0.0$ to $\unit[0.43]{\mu m}$ for $\mathcal{P}=0.6$ in the case of $T_*=\unit[7200]{K}$.
\item For particles with radii $\unit[0.03]{\mu m}<a< \unit[1.2]{\mu m}$ the following is valid. The higher the porosity, the colder is the particle. For a fixed radius, the drop in temperature decreases with increasing porosity. For $\mathcal{P} = 0.6$  and radii $a<\unit[0.03]{\mu m}$, the temperature is  even higher than for corresponding radii at $\mathcal{P}=0.5$.
\item The blowout sizes as calculated in Sect. \ref{Abschnitt38} are marked in the figures and reveal that most of the presented dust particles will be blown out of the system.
\end{enumerate}

The temperature of the blowout grains is an important parameter, because this is the quantity that is derived from observations. We calculated the dust temperature of the blowout grains $T_{\mathrm{BO}}$ at different distances from the star, since it might depend on the cooling efficiency, and hence on the stellar distance (Fig. \ref{TaBO}). The temperature of the blowout grains is at all distances for porous grains slightly lower than for compact spheres. For example, at $r=\unit[5]{AU}$ and for $\mathcal{P}=0.6$, the dust temperature is $\unit[22]{K}$ below the one for $\mathcal{P}=0.0$. At $\unit[100]{AU}$ this difference is $\unit[9.2]{K}$.

\cite{Vosh2} determined the dust temperature for porous particles  using the effective medium theory once and the method of layered spheres once. The material  was a mixture of  astronomical  silicate  (\citealt{DraineLaor1993}) and amorphous carbon (\citealt{Rouleau1991}) with a weighting of 1:1. Comparing our grain model  with  the studies from \cite{Vosh2}, we furthermore calculated  the dust temperature for a mixture of astronomical silicate and amorphous carbon, which is presented in Fig. \ref{TempAC1} as a function of the effective radius,  $a_{\mathrm{eff}}=(1-\mathcal{P})^{\nicefrac{1}{3}}\, a$. The central star has an effective temperature of $T_*=\unit[2500]{K}$, and the assumed grain distance is  $r = 10^4\,R_{*}$.

The results of \cite{Vosh2} are comparable to the ones calculated with the DDA. For the layered spheres the temperature decreases  with increasing porosity. This statement applies to the EMT only up to a porosity of $\mathcal{P} = 0.7 $, when the temperature increases again. The decrease in temperature for lower porosities can also be observed in our approach. \citet{Greenberg1990} ascertained in their studies the increasing of temperature at very high porosities, where the temperature approaches the value of subunits of the porous particle. Certainly the behavior differs for the lower porosities, since the temperature remains constant or increases slightly. \citet{Vosh2} have also performed investigations  for different stellar temperatures and distances  and found the same behavior. For high porosities, the differences between the method of layered spheres and the EMT could reach $\sim 15\, \%$, and  for compact grains they were less than $1 \,\%$.

In conclusion,  porous  particles with radii up to $\unit[1.3]{\mu m}$  are generally colder than compact spheres, and the maximum of the infrared emission is thus shifted to longer wavelengths. Our results agree very well with those of \citet{Vosh2}.

%#########################################################################################################################################
%#########################################################################################################################################
%#########################################################################################################################################
%###########################################            6        #########################################################################
%#########################################################################################################################################
%#########################################################################################################################################

\section{Conclusions}
When modeling the density and grain size distribution in debris disks, the minimum particle size is often significantly larger than the corresponding blowout size. 
We investigated porous dust grains, which provide a possible explanation. Since the introduced $\beta$-ratio depends on the optical properties of the grain, we determined the absorption cross sections of porous particles as a function of the porosity.  The blowout sizes  of six different sample stars and seven porosities (including compact spheres)  were calculated. We found that the blowout size increases with grain porosity and stellar temperature. It should be noted that both the choice of the bulk density, as well as the particle ejection limit, has a strong influence on the obtained results. An approximation equation was derived from the numerically calculated values with which the blowout size can be estimated. Finally, the dependence of the dust temperature on  particle radius and porosity was investigated. It was found that porous particles  are colder than compact spheres, and the temperature of blowout grains in particular decreases for porous particles.

The optical properties of porous particles were determined with the DDSCAT for an extended range of wavelengths and particle radii. These efficiency factors of absorption, scattering, and radiation pressure are summarized in several tables and are available for further applications. An online tool for calculating the blowout size based on the derived approximation function is available at  \textit{http://www1.astrophysik.uni-kiel.de/blowout/}.
\begin{figure}[b!]
	\centering
\includegraphics[width=1.0\linewidth]{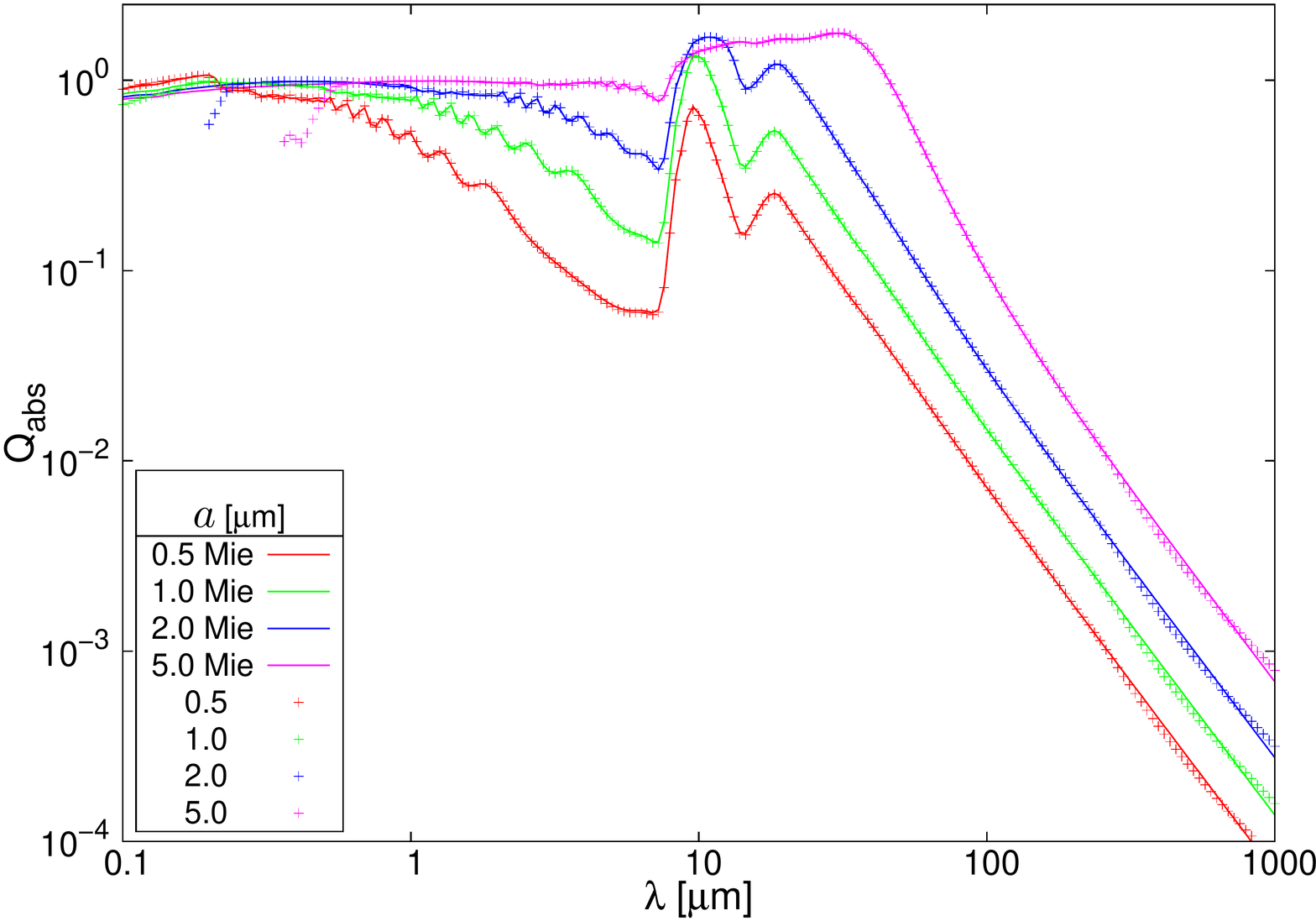}
\includegraphics[width=1.0\linewidth]{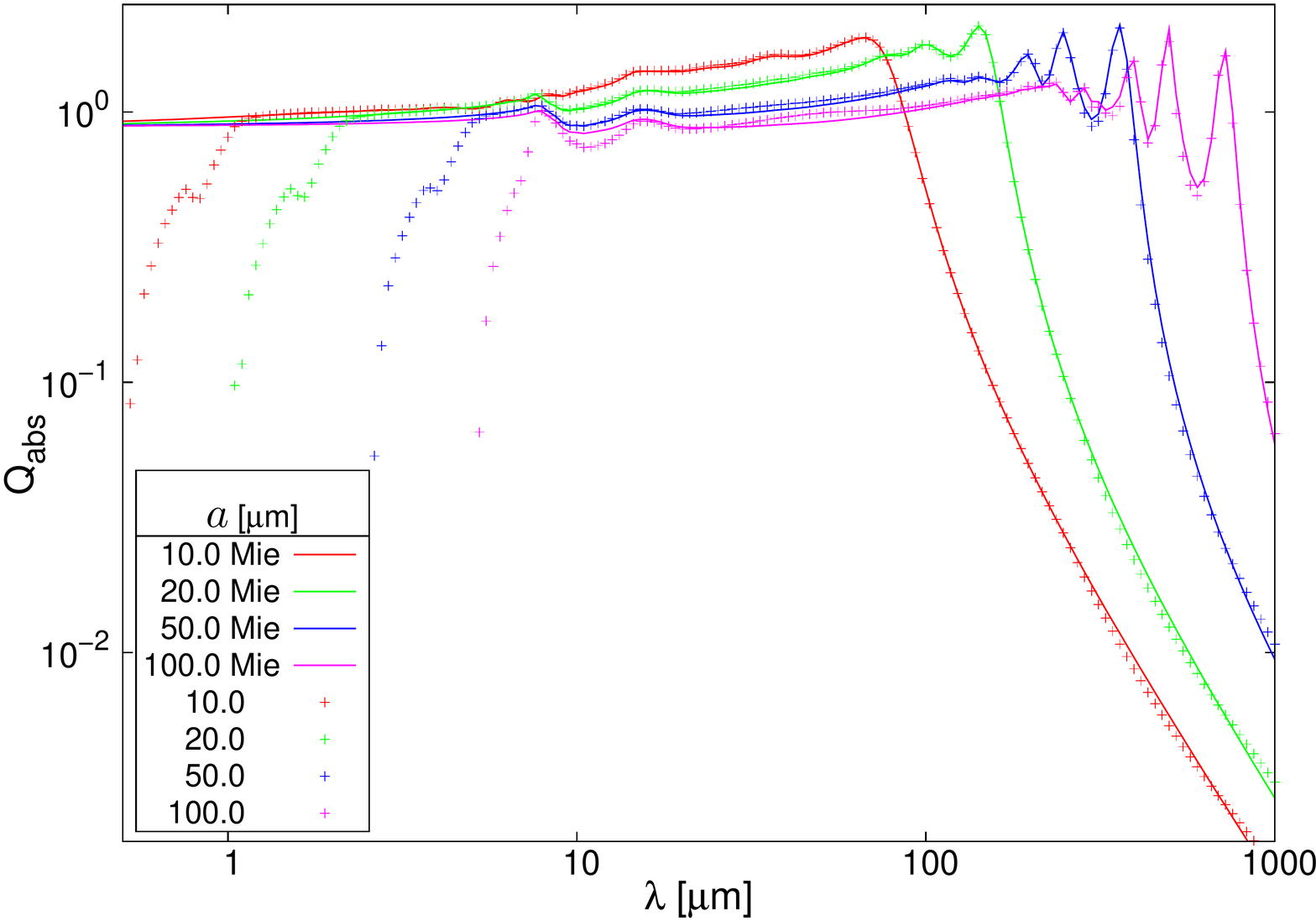}
	\caption{Absorption cross section  $Q_{\textup{abs}}$ derived with DDA (crosses) and Mie (solid lines) as a function of the wavelength. The particles are compact spheres  with radii $a$ from $0.5$ to $ \unit[5.0]{\mu  m}$ (above) and  $ 10.0 $ to $ \unit[100.0]{\mu m}$ (bottom).}
	\label{Vollkugeln}
\end{figure}
%__________________________________________________________________________________________________________________________________________________________________________
%__________________________________________________________________________________________________________________________________________________________________________
%__________________________________________________________________________________________________________________________________________________________________________
\begin{acknowledgements}
We would  like to  thank Steve Ertel for a helpful discussion. F.K. thanks the DFG for financial support under contract WO 857/7-1.
\end{acknowledgements}

%__________________________________________________________________________________________________________________________________________________________________________
%__________________________________________________________________________________________________________________________________________________________________________
%__________________________________________________________________________________________________________________________________________________________________________

\appendix
\setcounter{secnumdepth}{+2}
\setcounter{section}{0}

\renewcommand\thesection{\Alph{section}}
\section{Tests with compact grains}
\label{review}

\begin{figure}[b!]
	\centering
\includegraphics[width=1.0\linewidth]{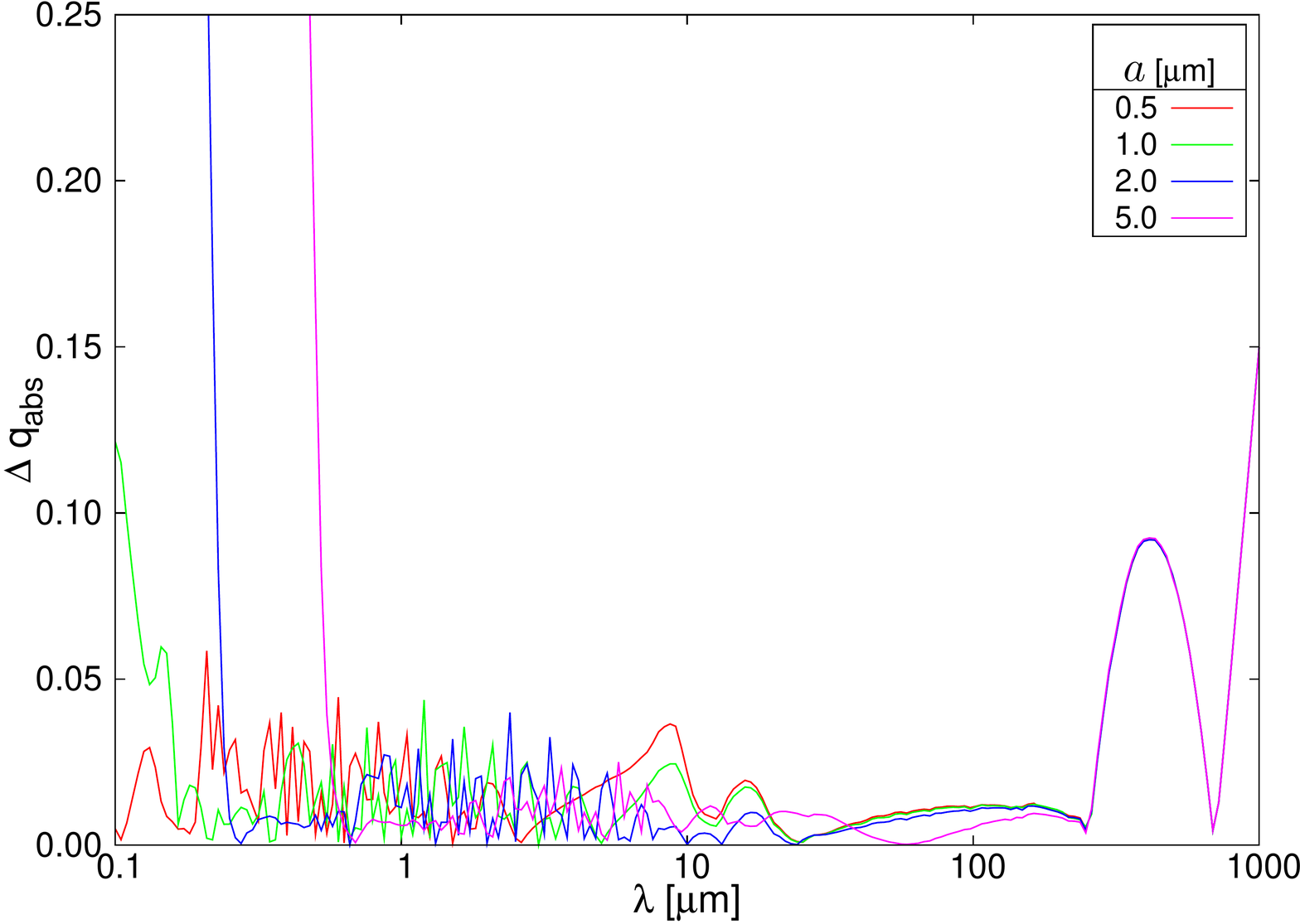}
\includegraphics[width=1.0\linewidth]{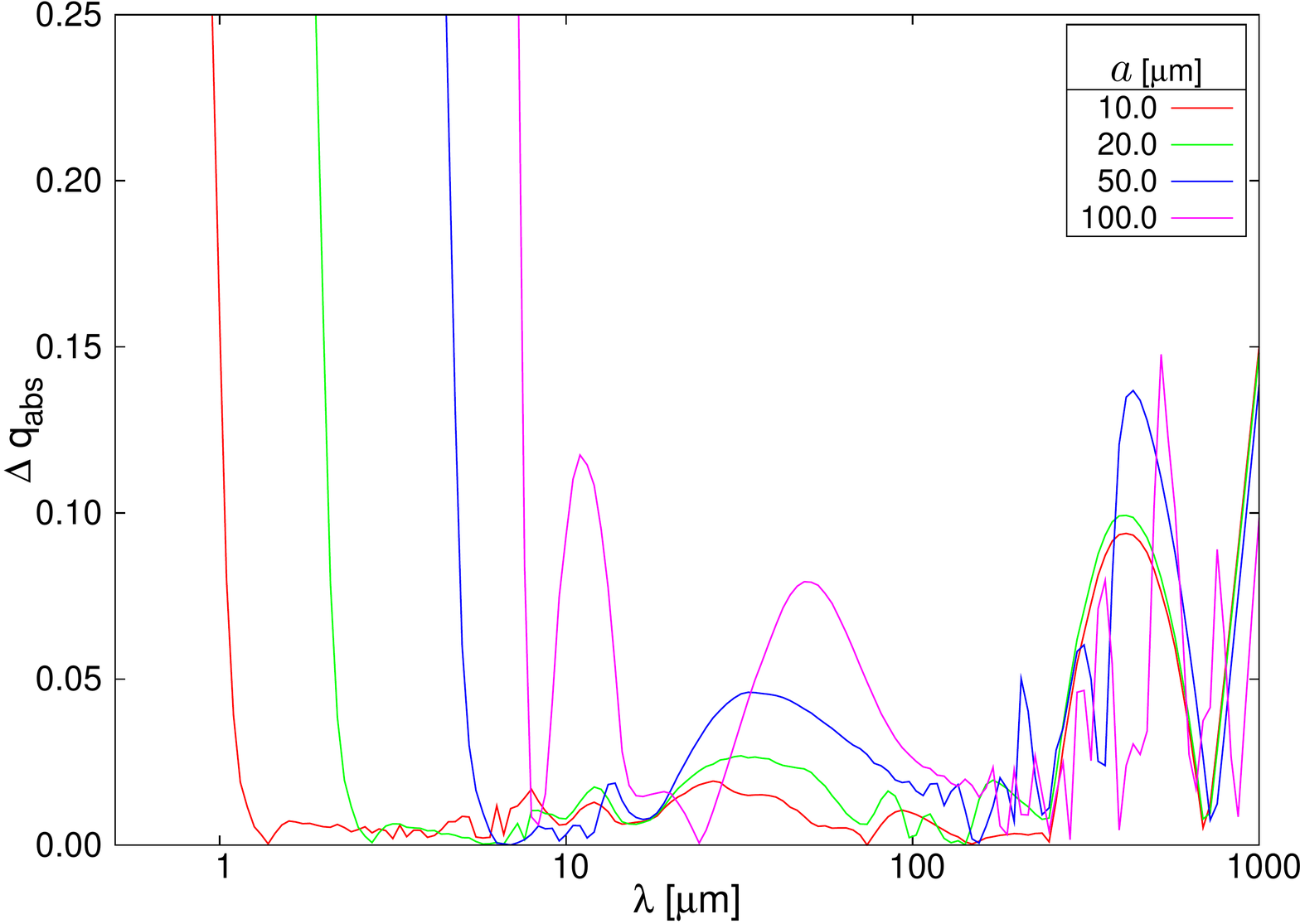}
	\caption{Relative deviations of the absorption cross sections derived with   DDA and Mie, $ \Delta q_{\textup{abs}} =  \arrowvert(Q_{\textup{abs}}-  Q_{\textup{abs,Mie}})/Q_{\textup{abs, Mie }}\arrowvert$.  The particles are  compact spheres with radii $a$ from $ 0.5 $ to $ \unit[5.0]{\mu  m}$  (above) and $ 10.0 $ to $\unit[100.0]{\mu  m} $ (bottom).}
	\label{relAbweichVollkugeln}
\end{figure}

\subsection{Comparison Mie theory and DDA}

We use the program \texttt{miex} (\citealt{WolfVoshchinnikov04}) to calculate the absorption cross sections based on Mie scattering and compare them with DDA results. Eight grain radii from $0.5$ to $\unit[100]{\mu m}$ and 200 wavelengths from $0.1$ to $\unit[1000]{\mu m}$ are examined (Fig. \ref{Vollkugeln}). With an increasing size parameter, the DDSCAT computing  time increases, and the results become highly inaccurate, depending on the chosen number of dipoles, in our case 519,832. In Fig. \ref{relAbweichVollkugeln} the relative deviations of the results of  DDA ($Q_{\textup{abs}}$) and Mie ($Q_{\textup{abs, Mie}}$)  are presented, $\Delta q_{\textup{abs}}=\break \arrowvert(Q_{\textup{abs}}-Q_{\textup{abs,Mie}})/Q_{\textup{abs,Mie}}\arrowvert$.
One can see in  Fig. \ref{Vollkugeln} that at short wavelengths the DDA values are significantly below the Mie ones. Thus these results  are neglected,  exceeding the limits of the  DDA, and there is a  lower limit of the wavelengths $\lambda_{\textup{min}}$ for which at all wavelengths $\lambda>\lambda_{\textup{min}}$ follows: $\Delta q_{\textup{abs}}(\lambda)<5\,\%$. Here, $\lambda_{\textup{min}}$ increases with the grain size $a$ (Table \ref{hiergehtslos}). The limit of applicability can be raised by increasing the number of dipoles, resulting in an increase in the calculation time as well.

In conclusion, to calculate absorption coefficients for wavelengths shorter than $\unit[0.5]{\mu m}$, the DDA can be applied within our parameter settings for grains with radii  $a \leq\unit[5]{\mu m}$. Despite these limitations, the similarities of both approaches are remarkably good.
\begin{table}[h!]
\centering
\vspace{3mm}\caption{The smallest wavelength $ \lambda_{\textup{min}} $, at which the relative deviations of the DDA and Mie results are not larger than $5\,\%$. The number of dipoles is 519,832.}\vspace{-3mm}
\begin{tabular}{|l|cccc|}\hline
$\hspace{0.2cm}a\hspace{0.15cm}\unit[]{[\mu m]}$      &$\phantom{i}0.5$&$1.0$&$2.0$&$5.0$\\
$\lambda_{\textup{min}}\unit[]{[\mu m]}$              &$<0.1$&$0.11$&$0.20$&$0.44$            \\\hline
$\hspace{0.2cm}a\hspace{0.15cm}\unit[]{[\mu m]}$     &$10.0$&$20.0$&$50.0$&$100.0$\\
 $\lambda_{\textup{min}}\unit[]{[\mu m]}$            &$0.87$&$1.7$ &$4.2$   &$\sim12$\\       \hline
\end{tabular}
		   \label{hiergehtslos}
\end{table}
\subsection{Decreased number of dipoles}
The convergence of DDA depends on the particle shape, and for compact spheres it is easier to reach convergence (\citealt{Yurkin2010}). To check the accuracy of the method we decrease the number of dipoles by a factor of 2 and compare the results. The decreased number of dipoles causes only small changes. The lower limit of the wavelengths  $\lambda_{\textup{min}}$ is slightly increased (Table \ref{hiergehtsweiter}). These values mark the upper limit of applicability of the DDSCAT (Fig. \ref{gultigBild}). Besides that, the changes are negligible.
\begin{table}[h!]
\centering
\vspace{3mm}\caption{Same as Table \ref{hiergehtslos}, but for the relative deviations between normal and halved number of dipoles.}\vspace{-3mm}
\begin{tabular}{|l|cccc|}\hline
$\hspace{0.2cm}a\hspace{0.15cm}\unit[]{[\mu m]}$      &$\phantom{i}0.5$&$1.0$&$2.0$&$5.0$\\
$\lambda_{\textup{min}}\unit[]{[\mu m]}$              &$<0.1$&$0.12$&$0.24$&$0.55$            \\\hline
$\hspace{0.2cm}a\hspace{0.15cm}\unit[]{[\mu m]}$     &$10.0$&$20.0$&$50.0$&$100.0$\\
 $\lambda_{\textup{min}}\unit[]{[\mu m]}$            &$1.08$&$2.2$ &$5.1$   &$\sim13$\\       \hline
\end{tabular}
		   \label{hiergehtsweiter}
\end{table}
\bibliography{aa}
\bibliographystyle{aa}

\end{document}